\documentclass[preprintnumbers,prd,twocolumn,aps,superscriptaddress,footinbib,amsfonts,amsmath,amssymb,nofootinbib,bm,showpacs,floatfix,showpreprintnumbers]{revtex4-1}

\usepackage[colorlinks=true,citecolor=blue,linkcolor=blue,urlcolor=blue]{hyperref}
\usepackage[utf8]{inputenc}
\usepackage{graphicx} 
\usepackage{epsfig}
\usepackage{epstopdf}
\usepackage{amsmath}
\usepackage{cancel}
\usepackage{amsfonts}
\usepackage{amssymb} 
\usepackage{multirow}
\usepackage{lineno}
\usepackage{url}
\usepackage{comment}
\usepackage{soul}
\usepackage[usenames,dvipsnames]{color}
\usepackage{balance}
\usepackage{csquotes}
\usepackage{textcomp} 
\graphicspath{{figures/}}
\usepackage{color,amsmath,amssymb,amsfonts,amsbsy}
\allowdisplaybreaks  
\usepackage{bbm}
\usepackage{bm}
\usepackage[dvipsnames]{xcolor}
\usepackage{relsize}
\usepackage{listings}
\usepackage{inconsolata}
\usepackage{sidecap}
\usepackage{subfigure}
\usepackage[normalem]{ulem}
\usepackage{lipsum}
\usepackage{graphicx} 
\usepackage{relsize}
\usepackage{tabularx}
\usepackage{dcolumn}
\usepackage{placeins} 
\usepackage{orcidlink}
\usepackage{lipsum}
\usepackage{footnote}
\usepackage[bottom]{footmisc}
\usepackage{enumitem}
\usepackage{physics}
\usepackage{slashed}

\def\ba{\begin{eqnarray}}
\def\ea{\end{eqnarray}}
\def\nn{\nonumber\\}
\def\ly{\lambda_Y}

\def\empile#1\over#2{\mathrel{\mathop{\kern 0pt#1}\limits_{#2}}}

\def\beq{\begin{equation}}
\def\eeq{\end{equation}}
\def\bea{\begin{eqnarray}}
\def\eea{\end{eqnarray}}

\def\p{{\boldsymbol p}}

\def\d3p{\frac{d^3\p}{(2\pi)^3}E_\p}

\catcode`\@=11

\newcount\@tempcntc
\def\@citex[#1]#2{\if@filesw\immediate\write\@auxout{\string\citation{#2}}\fi
  \@tempcnta\z@\@tempcntb\m@ne\def\@citea{}\@cite{%
        \@for\@citeb:=#2\do%
    {\@ifundefined{b@\@citeb}%
        {\@citeo\@tempcntb\m@ne\@citea%
                \def\@citea{,\penalty\@m\ }{\bf ?}\@warning%
                {Citation `\@citeb' on page \thepage \space undefined}}%
        {\setbox\z@\hbox{\global\@tempcntc0\csname b@\@citeb\endcsname\relax}
     \ifnum\@tempcntc=\z@ \@citeo\@tempcntb\m@ne%
       \@citea\def\@citea{,\penalty\@m}%
       \hbox{\csname b@\@citeb\endcsname}%
     \else%
      \advance\@tempcntb\@ne%
      \ifnum\@tempcntb=\@tempcntc%
      \else\advance\@tempcntb\m@ne\@citeo%
      \@tempcnta\@tempcntc\@tempcntb\@tempcntc\fi\fi}}\@citeo}{#1}}%

\def\@citeo{\ifnum\@tempcnta>\@tempcntb\else\@citea
  \def\@citea{,\penalty\@m}%
  \ifnum\@tempcnta=\@tempcntb\the\@tempcnta\else
   {\advance\@tempcnta\@ne\ifnum\@tempcnta=\@tempcntb \else
\def\@citea{--}\fi
    \advance\@tempcnta\m@ne\the\@tempcnta\@citea\the\@tempcntb}\fi\fi}

\catcode`\@=12

\begin{document}

\newcommand*{\DUKE}{Physics Department, Duke University, Durham, North Carolina 27708, USA
}\affiliation{\DUKE}
\newcommand*{\TRI}{Triangle Universities Nuclear Laboratory, Durham, North Carolina 27708, USA
}\affiliation{\TRI}
\newcommand*{\BSU}{Institute for Nuclear Problems of Belorussian State University, Minsk, 220006, Belarus
}\affiliation{\BSU}
\newcommand*{\MIS}{Department of Physics and Astronomy, Mississippi State University, Starkville, Mississippi 39762, USA
}\affiliation{\MIS}
\newcommand*{\IND}{Physics Department, Indiana University, Bloomington, Indiana 47405, USA
}\affiliation{\IND}
\newcommand*{\SU}{Key Laboratory of Particle Physics and Particle Irradiation (MOE), Institute of Frontier and Interdisciplinary Science, Shandong University, Qingdao, Shandong 266237, China
}\affiliation{\SU}
\newcommand*{\IMP}{Southern Center for Nuclear Science Theory, Institute of Modern Physics, Chinese Academy of Sciences, Huizhou, 516000, China
}\affiliation{\IMP}
\newcommand*{\JLAB}{Theory Center, Jefferson Lab, Newport News, Virginia 23606, USA}

\preprint{JLAB-THY-26-4907}

\title{QED radiative effects in semi-inclusive deep-inelastic scattering: \\ traditional and factorized approaches}

\author{Igor Akushevich}
\affiliation{\DUKE}
\author{Haiyan Gao}
\affiliation{\DUKE}
\affiliation{\TRI}
\author{Alexander Ilyichev}
\affiliation{\BSU}
\author{Shuo Jia}
\affiliation{\DUKE}
\author{Vladimir~Khachatryan}\email{vk355@msstate.edu} 
\affiliation{\DUKE}
\affiliation{\TRI}
\affiliation{\MIS}
\affiliation{\IND}
\author{Youjie Lin}
\affiliation{\SU}
\author{Tianbo Liu}
\affiliation{\SU}
\affiliation{\IMP}
\author{W.~Melnitchouk}
\affiliation{\JLAB}

\date{\today}

\begin{abstract}
Semi-inclusive deep-inelastic scattering (SIDIS) of leptons is a vital tool for probing the three-dimensional momentum space partonic structure of the nucleon. 
Reliable extraction of the intrinsic collinear or transverse momentum dependent parton distributions from SIDIS data requires careful treatment of QED radiative effects beyond the Born level.
In this work we perform a detailed comparative analysis of QED effects in SIDIS using the traditional Bardin-Shumeiko approach, augmented by the electron structure function method, and a more recent factorized approach that treats QED and QCD radiation on equal footing. 
We compare analytical and numerical results of these approaches for the unpolarized $\pi^+$ electroproduction cross sections off the proton and the Collins and Sivers transverse single-spin asymmetries, and identify the sources of discrepancies between the calculations, which for the cross sections can reach up to $\approx 2\%$, $\approx 7\%$, and $\approx 5\%$ at Jefferson Lab, HERMES, and EIC kinematics, respectively.
To provide a unified approach for reliably extracting partonic information from future SIDIS measurements, we propose a hybrid framework that employs the salient features of both frameworks in their respective regions of applicability.
\end{abstract}
\maketitle

\section{Introduction \label{eq::intro}}

Exploration of the internal structure of hadrons in terms of their fundamental quark and gluon (or parton) building blocks has been an intriguing and enduring scientific quest for over half a century. 
In this regard, the nature of the phenomenon of quark confinement remains not fully understood within quantum chromodynamics (QCD), the theory of the strong nuclear interactions, despite considerable progress over the years in studying its nonperturbative features~\cite{Gross:2022hyw}.

One avenue for investigating confinement is through the study of nucleon structure in an effort to understand the motion of confined partons inside the proton, as well as the distribution of the spin among the proton's quark and gluon constituents. 
Dedicated ongoing and future research efforts, both experimental and theoretical, aim to unravel the nucleon structure in both the momentum and spatial dimensions. 
Several decades of high-energy scattering experiments, including deep-inelastic scattering (DIS), Drell-Yan lepton-pair production, and inclusive weak boson or jet production in hadronic collisions, has led to considerable accumulated knowledge about the longitudinal momentum distributions of quarks and gluons inside the nucleon, parametrized through parton distribution functions (PDFs)~\cite{Jimenez-Delgado:2013sma, Alekhin:2017kpj, Ethier:2020way, Bailey:2020ooq, NNPDF:2021njg, Cocuzza:2025qvf, Cerutti:2025yji, Cocuzza:2026zoy}.

More recently, efforts to expand our knowledge of nucleon structure into the transverse dimension have led to the development of new theoretical frameworks based on transverse momentum dependent (TMD) factorization~\cite{Collins:2011zzd, Boussarie:2023izj}.
These frameworks have allowed the extraction of TMD~PDFs, from which the 3-dimensional structure of the nucleon can be built up through nucleon tomography in momentum space~\cite{Bacchetta:2006tn, geles_Martinez_2015, Diehl:2015uka}. 
The study of TMD~PDFs also gives insight into the quark and gluon orbital angular momentum in the proton, which remains the main missing component in the proton spin puzzle~\cite{STAR, HERMES, COMPASS, PHENIX:2007kqm, Ji_review, KUHN20091}.

The simplest process which can be used to study TMD~PDFs is semi-inclusive DIS (SIDIS) of charged leptons from nucleons, with the production of a fast hadron (typically pion) in coincidence with the scattered lepton~\cite{ANSELMINO2020103806}. 
Leptons are naturally cleaner probes of nucleon structure, since the internal structure of a hadron beam is inevitably entangled with that of the nucleon that is being probed.
It is well known, however, that charged lepton scattering can generate electromagnetic radiation that can significantly alter the apparent nucleon structure that would otherwise be extracted in the absence of radiation, especially at high momentum transfers.

The importance of understanding radiative corrections (RCs) at the sub-percent level for elastic lepton scattering~\cite{Mo_Tsai, Maximon:2000hm} was dramatically illustrated by the extraction of the proton's electric to magnetic form factor ratio from the Rosenbluth and polarization transfer methods, which highlighted the critical role played by two-photon exchange (TPE) contributions~\cite{Guichon:2003qm, Blunden:2003sp, Carlson:2007sp, Arrington:2011dn, Afanasev:2017gsk}. 
Similarly, for inclusive scattering, QED radiation can give rise to large corrections, especially at small values of Bjorken-$x$ and large lepton inelasticity, corresponding to regions with more phase space for radiation~\cite{Spiesberger:1990fa, Badelek:1994uq, Akushevich:1994dn, Blumlein:2002fy}.
For SIDIS reactions, with more kinematic variables and correlations between them, the radiative effects can be even more severe than for inclusive DIS.
In this case, the QED radiation can modify the angular modulations between the leptonic scattering and hadronic production planes, and the radiation-induced angular modulations can imitate those stemming from the nucleon structure effects containing the sought-after TMD PDFs~\cite{Liu:2020rvc, Liu:2021jfp}.
It is vital, therefore, for the reliable extraction of structural information on the nucleon from SIDIS processes to account for QED RC effects in a systematic and robust way.

In this paper we present a comprehensive analytical and numerical analysis of RC effects in SIDIS using different theoretical approaches developed recently~\cite{Akushevich:2019mbz, Liu:2020rvc, Liu:2021jfp, Tra_RC_exc, Akushevich:2024uhb}. 
These include the traditional Bardin-Shumeiko \cite{Akushevich:2019mbz, Tra_RC_exc} and electron structure function \cite{Akushevich:2024uhb} approaches,%
\footnote{The work of Ref.~\cite{Akushevich:2019mbz} follows and evolves from the original paper by Mo and Tsai~\cite{Mo_Tsai} that treats QED effects as corrections.}
and the QED+QCD factorization scheme \cite{Liu:2020rvc, Liu:2021jfp}.
In practice, the numerical results from these calculations can differ by amounts greater than the desired precision of ongoing and planned SIDIS programs at Jefferson Lab (JLab), such as those involving the Solenoidal Large Intensity Device (SoLID)~\cite{289ae0, SoLID_white}, and the future Electron-Ion Collider (EIC)~\cite{Accardi:2012qut, AbdulKhalek:2021gbh}, where the RC effects are expected to become more important with the greater phase space  available to shower compared to lower center-of-mass energy facilities.%
\footnote{The ansatz developed in Ref.~\cite{Akushevich:1998dz} for one-loop electroweak RCs to the lepton current in DIS for scattering longitudinally polarized leptons from polarized nucleons can also be extended to SIDIS (using the framework of Refs.~\cite{Akushevich:2019mbz, Tra_RC_exc, Akushevich:2024uhb}) and applied to EIC kinematics.}
It is timely, therefore, to benchmark the methods in order to systematically assess their merits and improve the quantitative understanding of RCs across specific contributions within different theoretical approaches and approximations used, with the goal of determining systematic uncertainties due to RCs for SIDIS experiments.
In particular, we will focus on quantifying the systematic uncertainties associated with RCs for the unpolarized differential SIDIS cross section, and for the Collins and Sivers transverse single-spin asymmetries (SSAs) in the the scattering of an unpolarized lepton beam off a transversely polarized proton target.

The rest of the paper is organized as follows.
In Sec.~\ref{eq::sidis} we introduce the basic elements of the SIDIS process, including definitions of kinematic variables and expressions for the differential cross sections in terms of the complete set of semi-inclusive structure functions. 
In Sec.~\ref{eq::RCs_gen} we discuss how the RCs are treated in SIDIS, rendering also some historical context. 
We present detailed formalisms of the Bardin-Shumeiko, electron structure function, and factorized approaches in Sec.~\ref{eq::RCs_frameworks}, demonstrating self-contained compilations of the theoretical results, before discussing numerical outputs in Sec.~\ref{eq::RCs_res} for the cases of the SIDIS unpolarized cross section and transverse SSAs at the JLab, HERMES, and EIC (at low center-of-mass energies) kinematics. 
In Sec.~\ref{eq::dis} we discuss the results, and in Sec.~\ref{eq::sum} give some conclusions along with prospects for future developments.

\section{Basics of SIDIS \label{eq::sidis}}

\begin{figure}[b]
\centering
\includegraphics[scale=0.625]{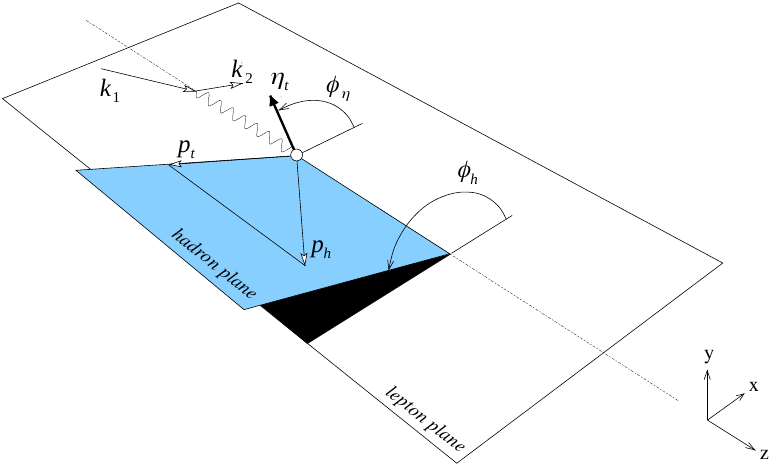}
\caption{Schematic view of the SIDIS process for the production of a hadron with four-momentum $p_h$ in the target rest frame~\cite{PhysRevD.70.117504}, with $p_t$ and $\eta_t$ the transverse parts of $p_h$ and target polarization $\eta$ with respect to the virtual photon momentum, with $\phi_h$ and $\phi_\eta$ the respective azimuthal angles.}
\label{SIDIS_Kinematics}
\end{figure}

In the SIDIS process, a lepton $\ell$ scatters from a nucleon $N$ with a final-state hadron $h$ detected in coincidence with the scattered lepton $\ell^\prime$ (see Fig.~\ref{SIDIS_Kinematics}),
\begin{equation}
\ell(k_1,\xi_{b}) + N(p,\eta)\ \rightarrow\ \ell^\prime(k_2) + h(p_h) + X(p_X) ,
\label{eq::SIDIS}
\end{equation}
with $X$ denoting undetected final-state particles.
Here, $k_1$ and $p$ ($k_2$ and $p_h$) are the four-momenta of the initial state (final state) leptons and hadrons (with invariant squared masses $k_1^2=k_2^2=m^2$, $p^2=M^2$, and $p_h^2=m_h^2$), and $\xi_{b}$ and $\eta$ are the incident beam and target nucleon polarization vectors, respectively.
The four-momentum squared of the final-state particles is bounded below by threshold pion production,
\ba
p_X^2 = (p+q-p_h)^2 > M_{\rm th}^2 \equiv (M+m_\pi)^2,
\label{thresh}
\ea 
where $m_\pi$ is the pion mass.
We also denote the four-momentum transfer by \mbox{$q = k_1 - k_2$}, with $Q^2 \equiv -q^2$.

In general, the polarized SIDIS process can be described by a set of six kinematic variables $\{ x, y, z, p_t, \phi_h, \phi \}$, three of which are the Lorentz invariant Bjorken scaling variable $x$, lepton inelasticity $y$, and hadron fragmentation variable $z$, where
\ba
x = \frac{Q^2}{2p \cdot q}, \qquad
y = \frac{p \cdot q}{p \cdot k_1}, \qquad
z = \frac{p \cdot p_h}{p \cdot q}.
\ea
The three Lorentz non-invariant variables are defined in the target rest frame, as illustrated in Fig.~\ref{SIDIS_Kinematics}: 
$p_t$ is the transverse component of the three-momentum $\bm{p}_h$ of the produced hadron $h$ with respect to the virtual photon three-momentum $\bm{q}$; 
$\phi_h$ is the angle between the leptonic and hadronic planes, defined according to the Trento Convention \cite{PhysRevD.70.117504}; and
$\phi$ is the azimuthal angle of the scattered lepton $\bm{k}_2$ around the lepton beam axis with respect to an arbitrary fixed direction. 
In the case of a transversely polarized target, the azimuthal angle $\phi$ is chosen to be the direction of $\eta$, $\dd\phi \approx \dd\phi_\eta$.

\begin{figure}[t]
\scalebox{0.55}{\includegraphics{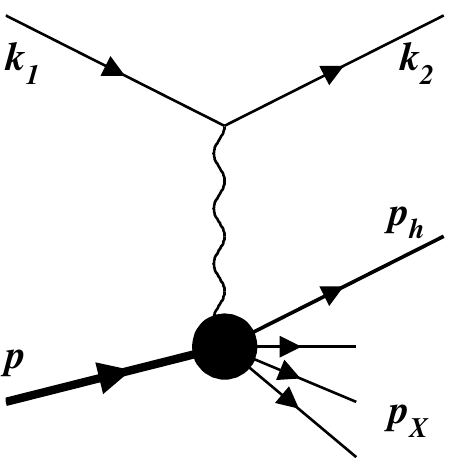}}
\caption{One-photon exchange (Born) approximation to the SIDIS process,  
$\ell(k_1)+N(p) \to \ell^\prime(k_2)+h(p_h)+X(p_X)$, with the particle momenta indicated on the external lines.}
\label{fgb}
\end{figure}

In the one-photon exchange (Born) approximation, depicted in Fig.~\ref{fgb}, the sixfold differential cross section can be written as a product of the leptonic ($L_B^{\mu \nu}$) and hadronic 
($W_{\mu\nu}$) tensors:
\begin{eqnarray}
\dd\sigma_B=\frac{(4\pi \alpha)^2}{2\sqrt{\lambda_S}\, Q^4} 
L_B^{\mu \nu}\, W_{\mu \nu}\,\dd\Gamma_B ,
\label{eq:sigma_B_had_lep}
\end{eqnarray}
where $\lambda_S=S^2-4 m^2 M^2$ and $S=2 p \cdot k_1$. 
The phase space $\dd\Gamma_B$ depends on the six kinematic variables that characterizes the final state according to
\begin{eqnarray}
\dd\Gamma_B 
&=& (2\pi)^4
\frac{\dd^3 k_2}{(2\pi)^3\, 2k_2^0}
\frac{\dd^3 p_h}{(2\pi)^3\, 2p_h^0}
\nonumber\\
&=& \frac{1}{4(2\pi)^2}
\frac{S\, S_x\, \dd x\, \dd y\, \dd\phi}{2\sqrt{\lambda_S}}\,
\frac{S_x\, \dd z\, \dd p_t^2\, \dd\phi_h}{4Mp_l},
\end{eqnarray}
where $k_2^0$ and $p_h^0$ are the final state lepton and hadron energies, respectively, $p_l=\sqrt{(p_h^0)^2-p_t^2-m_h^2}$ is the longitudinal momentum of the scattered lepton, and \mbox{$S_x=2p \cdot q$}.

The leptonic tensor is well known and can be found, for example, in Eq.~(11) of Ref.~\cite{Akushevich:2019mbz}. 
From current conservation, hermiticity, and parity conservation, the hadronic tensor for the amplitude $\gamma^* + n \to h + X$ can be decomposed into 5 scalar spin-independent and 13 spin-dependent semi-inclusive structure functions~\cite{1995}.
Using the notation from Ref.~\cite{Bacchetta:2006tn}, the SIDIS six-fold differential cross section, including all possible beam and target polarizations, can be written in the massless lepton limit ($m \to 0$) as
\begin{widetext}
\begin{eqnarray}
\label{born_xsection2}
\frac{\dd^6\sigma_B}{\dd x\, \dd y\, \dd\phi\, \dd z\, \dd\phi_h\, \dd p_t^2}
\!\!&=&\!\! \frac{\alpha^2}{x y\, Q^2}\,
\frac{y^2}{2\,(1-\varepsilon)}\,  \biggl( 1+\frac{\gamma^2}{2x} \biggr)\,
\Biggl\{
  F_{UU ,T} 
+ \varepsilon F_{UU ,L}
+ \sqrt{2\,\varepsilon (1+\varepsilon)}\,\cos\phi_h\, 
  F_{UU}^{\cos\phi_h}
+ \varepsilon \cos(2\phi_h)\, 
  F_{UU}^{\cos 2\phi_h}
\nonumber
\\
& & \hspace*{-2.8cm}
+\ \lambda_e\, \sqrt{2\,\varepsilon (1-\varepsilon)}\, \sin\phi_h\,
  F_{LU}^{\sin\phi_h}
+ \eta_L\, 
\bigg[ 
  \sqrt{2\, \varepsilon (1+\varepsilon)}\, \sin\phi_h\, 
  F_{UL}^{\sin\phi_h}
+ \varepsilon\, \sin(2\phi_h)\, 
  F_{UL}^{\sin 2\phi_h}
\bigg]
\nonumber
\\
& & \hspace*{-2.8cm}
+\ \eta_L \lambda_e\, 
\bigg[
  \sqrt{1-\varepsilon^2}\; 
  F_{LL}
+ \sqrt{2\,\varepsilon (1-\varepsilon)}\, \cos\phi_h\, 
  F_{LL}^{\cos \phi_h}
\bigg]
\nonumber
\\
& & \hspace*{-2.8cm}
+\ \eta_t\, 
\bigg[
  \varepsilon\, \sin(\phi_h+\phi_\eta)\, 
  F_{UT}^{\sin (\phi_h +\phi_\eta )}
+ \sin(\phi_h-\phi_\eta)\,
  \Bigl( F_{UT ,T}^{\sin(\phi_h -\phi_\eta )}
         + \varepsilon\, F_{UT ,L}^{\sin (\phi_h -\phi_\eta )} 
  \Bigr)
\nonumber
\\ 
& & \hspace*{-1.8cm}
+\ \varepsilon\, \sin(3\phi_h-\phi_\eta)\,
  F_{UT}^{\sin (3\phi_h -\phi_\eta )}
+ \sqrt{2\,\varepsilon (1+\varepsilon)}\, 
  \Big( \sin\phi_\eta\, F_{UT}^{\sin \phi_\eta }
      + \sin(2\phi_h-\phi_\eta)\, F_{UT}^{\sin (2\phi_h-\phi_\eta)}
  \Big)
\bigg]
\nonumber
\\
& & \hspace*{-2.8cm}
+\ \eta_t\, \lambda_e\, 
\bigg[
  \sqrt{1-\varepsilon^2}\, \cos(\phi_h-\phi_\eta)\, 
  F_{LT}^{\cos(\phi_h -\phi_\eta)}
+ \sqrt{2\,\varepsilon (1-\varepsilon)}\, 
  \Big( 
    \cos\phi_\eta\, F_{LT}^{\cos\phi_\eta}
  + \cos(2\phi_h-\phi_\eta)\, F_{LT}^{\cos(2\phi_h - \phi_\eta)}
  \Big)
\bigg] 
\Biggr\} ,
\end{eqnarray}
\end{widetext}
where 
\ba
\varepsilon = \frac{1 - y - \frac14 \gamma^2 y^2}{1 - y + \frac12 y^2 + \frac14 \gamma^2 y^2}
\ea
is the ratio of longitudinal and transverse photon fluxes, $\gamma=2xM/\sqrt{Q^2}$, with the longitudinal/transverse components of the nucleon polarization vector $\eta$ to be related by $\eta_L^2 + \eta_t^2 = 1$.
The semi-inclusive structure functions in Eq.~(\ref{born_xsection2}) are defined such that the first and second subscripts indicate the beam and target polarizations, while the third subscript (where appropriate) specifies the virtual photon polarization (longitudinal or transverse).

\section{Historical perspective \label{eq::RCs_gen}}

The traditional paradigm for hadron structure studies in lepton-nucleon scattering isolates the contribution from the fundamental one-photon exchange between the lepton beam and nucleon target by computing and subtracting higher-order electromagnetic RCs from the experimental data.
The higher-order contributions include real photon radiation from the incoming and outgoing leptons and hadrons, as well as virtual particle effects, such as vertex corrections, vacuum polarization, and multi-photon exchange (or box diagrams).

Data analyses of DIS and SIDIS measurements typically employ RC procedures in which radiative effects are removed from the observed cross sections or asymmetries to extract the corresponding Born-level observables. 
In current practice, only the model-independent parts of the total RCs are removed explicitly, while parameterizations of the Born hadronic structure functions are taken from phenomenological fits to previous measurements and, when necessary, refined iteratively as part of the extraction procedure~\cite{Akushevich:1994dn, Byer:2022bqf}.
Historically, the RCs have been divided into model-independent and model-dependent contributions. 
The former comprise leptonic bremsstrahlung from the incoming and outgoing lepton legs, vacuum polarization by leptons and hadrons, and leptonic vertex corrections. 
The latter, in contrast, include photon radiation from hadrons, two-photon exchange, and other corrections involving hadronic electromagnetic structure beyond the one-photon exchange process.
The model-independent RCs can be expressed in terms of the hadronic structure functions, which in practice are obtained from fits to previous experimental measurements.
At the same time, the current practice to account for the model-dependent RCs is by implicit embedding in the fitted structure functions, thereby avoiding explicit assumptions about the underlying hadronic model in the evaluation of the model-independent RCs.
The ultimate goal to extract Born-level structural information of the hadron would require explicit treatment of model-dependent RCs.

The effects of multiple bremsstrahlung from the lepton line is also an important and largely underdeveloped part of the model-independent RCs.
At large momentum transfers relative to the lepton mass, $Q^2 \gg m^2$, these corrections give the largest contribution to the total RC, and can be computed to high accuracy in the ultrarelativistic approximation. 
In this case, the most general structure of the total cross section including radiation from leptons in QED can be written in the leading log approximation as
\ba
\sigma_{RC} 
= \sum\limits_{i=1}^\infty \alpha^i\sum\limits_{j=0}^i C_{ij}\, l_m^j
+ {\cal O}\left(\frac {m^2}{Q^2}\right),
\label{src11}
\ea 
where $l_m=\log(Q^2/m^2)$, and the coefficients $C_{ij}$ are independent of the lepton mass (in this paper the lepton mass will usually refer to the electron mass --- see Sec.~\ref{eq::RCs_frameworks}).
The terms with $C_{ii}$ and $C_{ii-1}$ in Eq.~(\ref{src11}) contain contributions proportional to $\alpha^i\, l_m^i$ and $\alpha^i\, l_m^{i-1}$, which we refer to as leading order (LO) and next-to-leading 
order (NLO) corrections, respectively.
At JLab kinematics (shown later in Fig.~\ref{fig:fig_unpoll_xs}), for example, the lowest-order coefficient of $l_m$ is of the order of $\sim 15$, which gives the dominant contribution to the cross section.
For the calculation of TPE and radiation of unobserved photons by hadrons, referred to as ``model-dependent'' RCs, it requires additional information about hadronic interactions, and thus introduces additional uncertainties.

Early calculations of photon radiation utilized the so-called ``soft-photon approximation'', in which the scattering amplitudes were approximated by their contributions at the photon pole, which was systematized and generalized to the case of multiphoton emission by Yennie, Frautschi and Suura~\cite{Yennie:1961ad}. 
In practical situations, however, it is not possible to completely suppress the photon radiation,%
\footnote{Here ``suppress" refers to finding a cut on the inelasticity that results in excluding events with a photon energy above the cut.}
and in analyses of experimental data it is necessary to take the detector geometry into account, especially for inclusive measurements.

The evaluation of RCs beyond the soft-photon approximation requires careful consideration of infrared divergences.
Mo and Tsai~\cite{Mo_Tsai} developed a framework for calculating RCs with radiation of hard photons from electrons in the one-loop approximation for elastic and inelastic $ep$ scattering, in which the infrared divergences could be isolated and canceled.
One of the limitations of the Mo and Tsai approach, however, is the approximate treatment of soft photon contributions, which leads to the dependence of the results on an artificial parameter that divides the phase space of the photon into soft and hard parts. 
In numerical evaluations, this parameter is chosen to be as small as possible to reduce the area that is estimated approximately, but not too small in order to avoid possible numerical instabilities.%
\footnote{Subsequently, Bardin and Shumeiko showed how this parameter cancels analytically in the cross section including RCs~\cite{Bardin:1976qa}.}
The treatment of the soft contributions was later improved by Maximon and Tjon~\cite{Maximon:2000hm}, who computed the soft photon emission exactly, but still neglected the momentum dependence in the numerators away from the soft photon limit.

The leading log approximation for the calculation of gluon radiation in QCD was suggested by Dokshitzer~\cite{Dsp}, Gribov and Lipatov~\cite{GLsp}, and Altarelli and Parisi~\cite{APsp}, while the application to the lepton current was demonstrated by De Rújula, Petronzio, and Savoy-Navarro~\cite{DeRujula}.
A QCD-based approach was adopted for real photon radiation at first order ${\cal O}(\alpha\, l_m)$ by Bl\"umlein~\cite{Blumlein}, to second order ${\cal O}\big( (\alpha\, l_m)^2 \big)$ by Kripfganz, Mohring and Spiesberger~\cite{Spies}, to third order ${\cal O}\big((\alpha\, l_m)^3 \big)$ by Skrzypek~\cite{Skrzypek}, as well as for the second-order subleading term ${\cal O}(\alpha^2\, l_m)$, and to fifth order ${\cal O}\big( (\alpha\, l_m)^5 \big)$ by Bl\"umlein and Kawamura~\cite{Blumlein2, Blumlein3}.
The complete resummation of the leading log terms to all orders in $\alpha$ was performed by Kuraev {\it et al.}~\cite{kuraev1}, who also showed how the subleading terms at all orders in $\alpha$ had to be accounted for in their resummation scheme \cite{kuraev2}.
Such a scheme was later applied to polarized DIS~\cite{AAM2004, ESFRAD} and initial-state QED radiation in $e^+ e^-$ collisions~\cite{Frixione:2019lga}. 

Based on the method developed by Bardin and Shumeiko~\cite{Bardin:1976qa}, the original formalism for RCs in SIDIS in the simple quark-parton model was suggested by Soroko and Shumeiko~\cite{SSh1, SSh2}, and later implemented in the POLRAD~2.0 FORTRAN code as a patch of the SIRAD program~\cite{Polrad}.
The results allowed the calculation of RCs for the 3-dimensional SIDIS cross section averaged over polar angles and transverse momentum of the final-state hadron.
The formalism was then generalized~\cite{ASSh} to allow the calculation of the five-dimensional SIDIS cross section in scattering of unpolarized particles, and the exclusive radiative tail (ERT) was first calculated in Ref.~\cite{2009}. 
The explicit expressions for RCs to SIDIS with initial state polarization were derived expressed in Ref.~\cite{Akushevich:2019mbz} and estimated numerically in Ref.~\cite{Tra_RC_exc}. 
Most recently, the RCs to unpolarized SIDIS in the leading log approximation were presented in Ref.~\cite{Akushevich:2024uhb}. 

In a parallel development, a new factorized approach to RCs was recently introduced~\cite{Liu:2020rvc, Liu:2021jfp} for inelastic lepton-hadron collisions, which simultaneously accounts for both QED and QCD radiation.
In other words, it is a theoretical framework in which both partonic and leptonic distributions can be determined consistently within a unified treatment of electromagnetic and strong interactions.
In analogy to the standard perturbative QCD factorization used for hard collisions, this approach separates the scattering process into hard parts that are calculable perturbatively, and nonperturbative soft parts that are absorbed into universal lepton distribution functions (LDFs) and lepton fragmentation functions (LFFs).
The hard QED radiation is treated perturbatively, together with the QCD radiation, leading to a tower of corrections of order $\mathcal{O}(\alpha^{m} \alpha_s^{n})$ to the Born amplitude. 
As demonstrated in this paper later, the factorized approach has a potential to provide a systematic framework for evaluating higher-order contributions to the model-independent component of RCs, thereby addressing one of the principal sources of uncertainty in SIDIS measurements where model-independent RCs are dominant, such as measurements of single-spin asymmetries. 
In addition to elaborate calculations of the model-independent RCs, another essential goal is to extract the Born-level hadronic structure functions from experimental data after explicit treatment of model-dependent RCs (as it was mentioned earlier in this section).
In this case, rather than applying RCs after the hadronic structure functions have been extracted, the factorized framework incorporates QED+QCD effects directly into the analysis of experimental data to extract structural information of the hadron at the Born level.
A strategy along these lines has recently been proposed for DIS~\cite{Cammarota:2025jyr}; however, a practical implementation suitable for an analysis of experimental data has yet to be developed for SIDIS.

\section{Radiative Corrections in SIDIS}
\label{eq::RCs_frameworks}

\begin{figure}[t]\centering
\scalebox{0.45}{\includegraphics{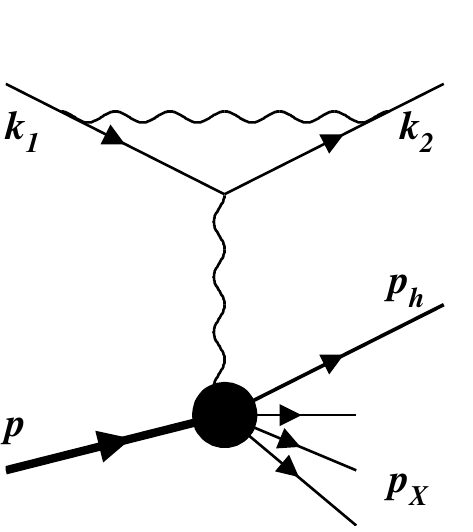}}
\scalebox{0.45}{\includegraphics{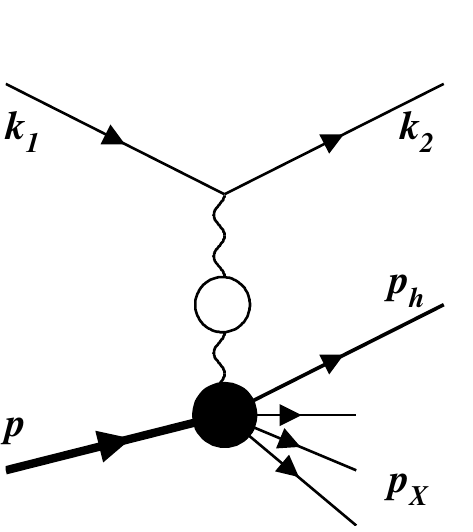}}
\\[-0.1cm]
{\bf \hspace{-.5cm} a) \hspace{3.3cm} b)\hspace{2.42cm}}
\\[0.1cm]
\scalebox{0.45}{\includegraphics{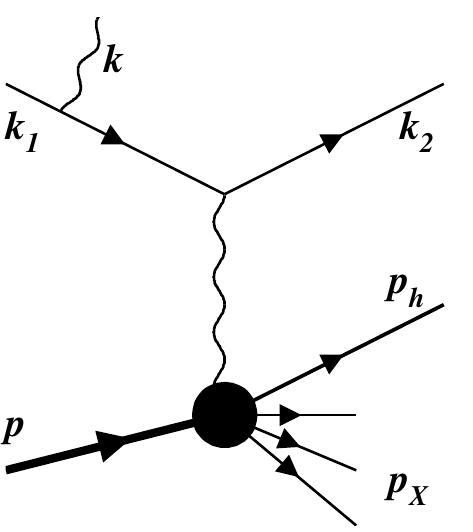}}
\scalebox{0.45}{\includegraphics{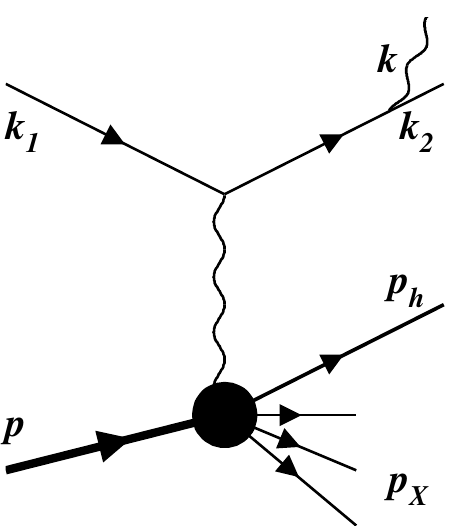}}
\\[-0.1cm]
{\bf \hspace{-.5cm} c) \hspace{3.3cm} d)\hspace{2.42cm}}
\\[0.1cm]
\scalebox{0.45}{\includegraphics{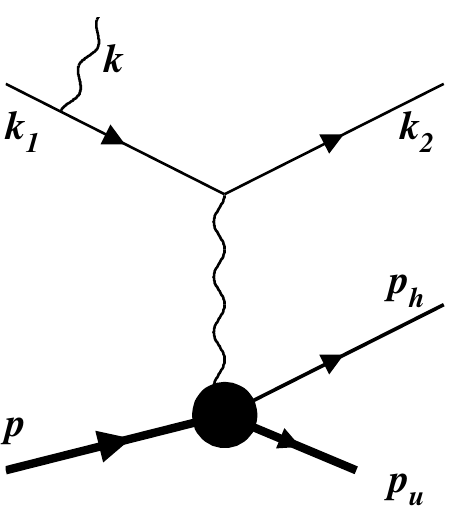}}
\scalebox{0.45}{\includegraphics{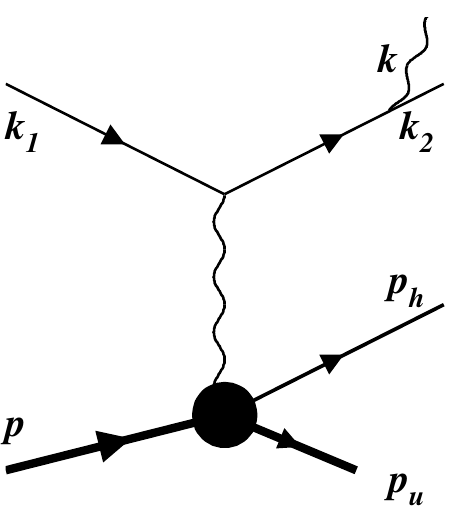}}
\\[-0.1cm]
{\bf \hspace{-.5cm} e) \hspace{3.3cm} f)\hspace{2.42cm}}
\\[0.1cm]
\caption{Feynman graphs for the contributions to the lowest-order RC in the semi-inclusive process [a)--d)], and the exclusive radiative tail (ERT) [e), f)]. The momenta of the initial and final-state particles are indicated, as in the text, and $p_u$ in diagrams e) and f) denotes the four-momentum of a single undetected hadron in the final state.}
\label{fgrc}
\end{figure}

Going beyond the Born-level amplitude for the SIDIS process depicted in Fig.~\ref{eq::SIDIS}, the set of one-loop corrections arising from photon radiation considered in the present analysis is illustrated in Fig.~\ref{fgrc}.
These include the leptonic vertex corrections [Fig.~\ref{fgrc}~a)], vacuum polarization [Fig.~\ref{fgrc}~b)], and bremsstrahlung, or unobserved real photon radiation, from the incident and outgoing leptons [Figs.~\ref{fgrc}~c)--d)].
The contributions from exclusive single hadroproduction, needed for the ERT, are also included [Figs.~\ref{fgrc}~e)--f)]. 
TPE contributions and radiation from hadrons, which are formally of the same order but depend also on hadron structure input, are not considered in this analysis (see Sec.~\ref{eq::dis}).

\subsection{Radiative effects with exact and leading-order (traditional) approaches \label{eq::RCs_lo}}

In this section, we summarize the essential results for the RCs to SIDIS within the traditional Bardin-Shumeiko and electron structure function methods.
Further details can be found in Refs.~\cite{2019, Tra_RC_exc, Akushevich:2024uhb}.

\subsubsection{Bardin-Shumeiko method}
\label{eq::RCs_lo1}

As discussed in Ref.~\cite{Akushevich:2019mbz}, the total radiation-included (or observed) SIDIS cross section has contributions from the Born term, $\sigma_B$ [Fig.~\ref{eq::SIDIS}], from semi-inclusive channels, $\sigma^{\rm in}$ [Figs.~\ref{fgrc}~a)--d)], and from the ERT, $\sigma^{\rm ex}_R$ [Figs.~\ref{fgrc}~e)--f)],
\ba
\sigma^{\rm obs}
= \sigma_B
+ \sigma^{\rm in}
+ \sigma^{\rm ex}_R,
\label{eq:siga_obs}
\ea
where ``$\sigma$'' is shorthand for the full, six-fold differential cross section as in Eq.~(\ref{born_xsection2}).
The contribution from the semi-inclusive final states is given by~\cite{Akushevich:2019mbz}
\begin{equation}
\sigma^{\rm in}
= \frac{\alpha }{\pi }
\big(
  \delta_{VR}
+ \delta_{\rm vac}^l
+ \delta_{\rm vac}^h
\big)\, \sigma_{B}
+ \sigma^F_R
+ {\sigma_{\rm AMM}} ,
\label{eq:siga_obs2}
\end{equation}
where $\delta_{\rm vac}^l$ and $\delta_{\rm vac}^h$ are the vacuum polarization contributions from leptons and hadrons, respectively, and {$\sigma_{\rm AMM}$} is the contribution from the anomalous magnetic moment~\cite{Akushevich:2019mbz}.
The factor $\delta_{VR}$ results from the cancellation of the infrared divergence, and $\sigma^F_R$ is the infrared-free contribution of the real photon radiation \cite{Akushevich:2019mbz}. 
The sum of the infrared divergent terms (in which the divergences cancel), including the soft photon radiation and vertex contribution, is given by
\newpage
\ba
\delta_{VR}
&=& 2 \big( (Q^2+2m^2) L_m-1 \big)\, 
    \log\frac{p_X^2-M_{\rm th}^2}{m \sqrt{p_X^2}}
\nn 
&& \hspace*{-1cm}
-\ \frac{Q^2+2m^2}{\sqrt{\lambda_m}}
\biggl[ 
  \frac12 \lambda_m L_m^2
+ 2{\rm Li}_2 
  \bigg( \frac{2\sqrt{\lambda_m}}{Q^2+\sqrt{\lambda_m}} \bigg)
- \frac{\pi^2}{2}
\biggr]
\nn 
&& \hspace*{-1cm}
+\ \frac12 S^\prime L_{S^\prime} + S_\phi - 2  
+  \frac 12 X^\prime L_{X^\prime}
+  \Big( \frac 32 Q^2 + 4 m^2 \Big) L_m,
\nn
&&
\label{dvr}
\ea 
where 
\begin{subequations}
\ba
L_m 
&=& \frac{1}{\sqrt{\lambda_m}}
    \log\frac{\sqrt{\lambda_m}+Q^2}{\sqrt{\lambda_m}-Q^2},
\\
L_{S^\prime}
&=& \frac 1{\sqrt{\lambda_S'}}
    \log\frac{S'+\sqrt{\lambda_S'}}{S'-\sqrt{\lambda_S'}},
\\
L_{X^\prime}
&=& \frac{1}{\sqrt{\lambda_X'}}
    \log\frac{X'+\sqrt{\lambda_X'}}{X'-\sqrt{\lambda_X'}}.
\label{lms}
\ea
\end{subequations} \\
Here, for completeness, the various kinematic variables are given by \cite{Akushevich:2019mbz}
\begin{subequations}
\ba
\lambda_m &=& Q^2 (Q^2 + 4 m^2),
\\
S^\prime &=& 2 k_1 \cdot p_X = S - Q^2 - V_1, 
\\
\lambda_S' &=& S'^2 - 4 m^2 p_X^2
\\
X^\prime &=& 2k_2 \cdot p_X = X + Q^2 - V_2, 
\\
\lambda_X' &=& X'^2 - 4 m^2 p_X^2
\\
X &=& 2 k_2 \cdot p,
\\
{V_{1,2}} &=& 2 k_{1,2} \cdot p_h,
\ea
\end{subequations}
and 
\begin{eqnarray}
{\rm Li}_2(x)=-\int\limits^x_0 \dd y\, \frac{\log|1-y|} y
\end{eqnarray}
is Spence's dilogarithm.
The explicit expression for $V_{1,2}$ and $S_\phi$ are given in Eq.~(5) of Ref.~\cite{Akushevich:2019mbz} and Eq.~(27) of Ref.~\cite{Tra_RC_exc}, respectively. \\

The finite part of real photon radiation in Fig.~\ref{fgrc}~c)--d) has the form
\ba
\sigma_R^F&=&-\frac{\alpha^3 S S_x^2}{64\pi^2 M p_l \lambda_S \sqrt{S_x^2+4M^2Q^2}
}
\nn[1mm]
&\times&
\int\limits_{\tau _{\rm min}}^{\tau _{\rm max}}
\dd\tau  
\int\limits_{0}^{2\pi}
\dd\phi_k  
\int\limits_{0}^{R _{\rm max}}
\dd R  
\nn[1mm]
&\times&
\sum\limits_{i=1}^9
\Biggl[
\sum\limits_{j=1}^{k_i}
\frac{{\cal H}_i(\tilde \chi_i^1,\tilde \chi_i^2,Q^2 + \tau R, \tilde x, \tilde z, \tilde p_t)\, \theta_{ij}\, R^{j-2}}{(Q^2+\tau R)^2}
\nn[1mm]
&&\qquad\qquad
-\
\frac{\theta_{i1}}R
\frac {{\cal H}_i(\chi_i^1,\chi_i^2,Q^2,x,z,p_t)}{Q^4}
\Biggr] ,
\label{srfin}
\ea
where the generalized structure functions ${\cal H}_{1-9}$ can be expressed through linear combinations of the 18 structure functions using Eq.~(19) of Ref.~\cite{Tra_RC_exc}.
In Eq.~(\ref{srfin}), the quantities $\chi_i^{1,2}$ are given by
    $$
    \chi_{1-5}^1 = 1,       \quad
    \chi_{6-9}^1 = \eta_1,  \quad
    \chi_{1-5}^2 = \eta_2,  \quad
    \chi_{6-9}^2 = \eta_3,  \quad
    $$
and the upper limits on the summations over $j$ are
    $k_i=\{3,3,3,3,3,4,4,4,4\}$.  
The components of the target nucleon spin vector $\eta$ in the generalized structure functions ${\cal H}_i$ are given by (see Ref.~\cite{Tra_RC_exc}):
\begin{subequations}
\label{etab}
\begin{eqnarray}
\eta_1 &=& \eta_t \cos(\phi_\eta-\phi_h),   \\
\eta_2 &=& \eta_t \sin(\phi_\eta-\phi_h),   \\
\eta_3 &=& -\eta_L.
\end{eqnarray}
\end{subequations}
The symbol\ ``$\sim$''\ over the arguments $x$, $z$, $p_t$ and $\eta_i$ in the structure functions ${\cal H}_i$ indicates that these are defined in terms of the shifted transfer momentum $q \to q-k$, where $k$ is the momentum of the radiated photon, as given in Eqs.~(35) and (A.1) of Ref.~\cite{Tra_RC_exc}.
The integration variables in Eq.~(\ref{srfin}) are given by
\ba
\tau = \frac{k \cdot q}{k \cdot p}, \qquad 
\phi_k, \qquad 
R = 2 k \cdot p, 
\ea
where $\phi_k$ is an angle between the $(\bm{k}_1, \bm{k}_2)$ and $(\bm{k}, \bm{q})$ planes. 
The limits of integration in Eq.~(\ref{srfin}) are given by
\begin{subequations}
\ba
R_{\rm max} &=& \frac{p_X^2 - M_{\rm th}^2}{1 + \tau - \mu},
\\
\tau_{\rm max/min} &=& \frac{S_x \pm \sqrt{S_x^2 + 4 M^2 Q^2}}{2M^2},
\ea
\end{subequations}
where $\mu = k \cdot p_h/k \cdot p $ is defined by Eq.~(34) of Ref.~\cite{Tra_RC_exc}. 
Explicit expression for the quantities $\theta_{ij}$ can be found in Appendix~B of Ref.~\cite{Akushevich:2019mbz}.

As for the SIDIS process, the exclusive reaction $\gamma^*  + N \to h + N'$ with the initial nucleon polarized can be described by 5 spin-independent and 13 spin-dependent structure functions.
For their representation, we again use 9 generalized exclusive structure functions that depend on the three variables $W^2 = (p+q)^2$, $Q^2$ and $t = (q-p_h)^2$, 
$${\mathcal H}^{\rm ex}_i(\chi_i^{1\,\rm ex}, \chi_i^{2\,\rm ex},W^2,Q^2,t),$$ 
where the quantities $\chi_i^{1,2 \rm ex}$ are given by
    $$
    \chi_{1-5}^{1\rm ex} = 1, \quad 
    \chi_{6-9}^{1\rm ex} = \eta^{\rm ex}_1, \quad
    \chi_{1-5}^{2\rm ex} = \eta^{\rm ex}_2, \quad
    \chi_{6-9}^{2\rm ex} = \eta^{\rm ex}_3. \quad
    $$
The components of the target polarization vector for the ERT, $\eta^{\rm ex}_{1-3}$, are obtained from the respective SIDIS quantities in Appendix A of Ref.~\cite{Tra_RC_exc}, with the replacements $R \to R_{\rm ex}$, 
where 
\ba
R_{\rm ex} &=& \frac{p_X^2-m_u^2}{1+\tau-\mu},
\ea
and $m_u$ is the mass of an undetected hadron.
With these definitions, the ERT\footnote{We also note that the Monte Carlo event generator for SIDIS RCs~\cite{Byer:2022bqf, SIDIS-RC_EvGen}, which is based on using the framework of Ref.~\cite{Akushevich:2019mbz}, needs to be updated to include the exclusive contribution in Eq.~(\ref{sre1}).} can be written as a double integral over $\tau$ and $\phi_k$~\cite{Akushevich:2019mbz},
\ba
\sigma^{\rm ex}_{R} = -\frac{\alpha^3\, S\, S_x^2}{2^9 \pi^5 M\, p_l\, \lambda_S \sqrt{S_x^2+4M^2Q^2}}
\int\limits_{\tau _{\rm min}}^{\tau _{\rm max}} \dd\tau
\int\limits_0^{2\pi}\dd\phi_k 
\nn
\sum_{i=1}^{9} \sum_{j=1}^{k_i}
\frac{{\mathcal H}^{\rm ex}_i(\tilde\chi_i^{1\rm ex},\tilde\chi_i^{2\rm ex}, Q^2+R_{\rm ex}\tau,{\widetilde W}^2, \tilde t)\, 
\theta_{ij}\, R_{\rm ex}^{j-2}}{(1+\tau-\mu)(Q^2+R_{\rm ex}\tau)^2},
\label{sre1}
\ea
where
\begin{subequations}
\ba
\widetilde W^2 &= W^2 - (1+\tau)\, R_{\rm ex},
\\
\tilde t &= t + (\mu-\tau)\, R_{\rm ex}.
\ea
\end{subequations}

\subsubsection{Leading-log method\label{eq::RCs_lo2}}

\begin{figure}[t]
\includegraphics[width=8.0cm,height=6.75cm]{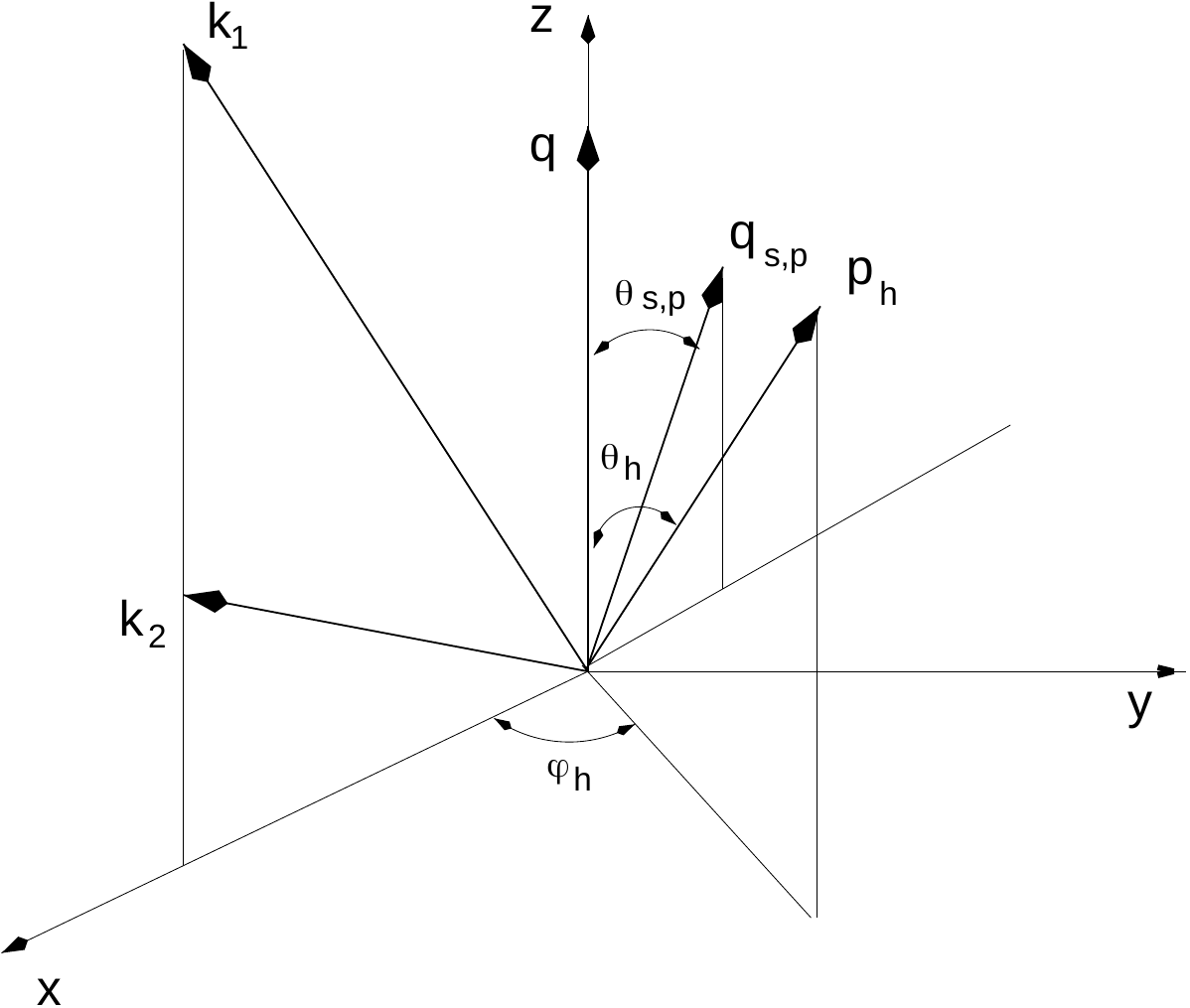}
\caption{Particle momenta in the SIDIS process (\ref{eq::SIDIS}) in the target rest frame; $q$ and $q_{s,p}$ are the momenta of the virtual photon in the original and shifted kinematics of the $s$- and $p$-peaks. }
\label{vect}
\end{figure}

In the standard leading-log approximation, the collinear real photon radiation along the initial and final electron directions gives rise to the so-called $s$- and $p$-peaks, respectively, whose contributions are proportional to $l_m$~\cite{Mo_Tsai}.
The corresponding four-vector $k^\mu$ of the radiated photon in the $s$- and $p$-peaks is parametrized by introducing the dimensionless variables $z_1$ and $z_2$ that reflect the remaining photon energy degree of freedom,
\begin{subequations}
\label{kpar}
\begin{align}
k_s &= (1-z_1)\, k_1,
\\
k_p &= (z_2^{-1}-1)\, k_2,
\end{align}
\end{subequations}
for the $s$- and $p$-peaks, respectively. 
This strategy has a useful geometric interpretation, allowing one to write the collinear bremsstrahlung in terms of the Born cross section in the so-called ``shifted Born condition''~\cite{Akushevich:2024uhb}, as shown in Fig.~\ref{vect},
\begin{subequations}
\ba
z_1 k_1 + p &=& k_2 + p_h + p_X,
\\
k_1 + p &=& \frac{1}{z_2} k_2 + p_h + p_X,
\label{kpar1}
\ea
\end{subequations}
where $q$ is chosen along the $z$-axis, and the lepton vectors $k_1$ and $k_2$ are in the $xz$-plane. 
This fixes the coordinate system. 
In the leading approximation, $k \to (1-z_1)\, k_1$ or $k \to (z_2^{-1}-1)\, k_2$ lies entirely in the $xz$-plane.

Applying this method to the calculation of RCs from the continuum spectrum, we find that the LO RCs in the first order in respect of $\alpha$ is
\ba
\sigma^{\rm in}_{{1L}}
&=&\Big[1 + \frac{\alpha}{\pi}\, \delta_{\rm vac}^l(Q^2)\Big]\, \sigma_B(S,Q^2,x,z,p_t,\phi_h,\eta_L,\phi_\eta)
\nonumber\\
&& +\ \sigma^{{\rm in}\;s}_{1L} + \sigma^{{\rm in}\;p}_{1L} ,
\label{xs_contin}
\ea
where
\begin{subequations}
\label{eq:dsigin1L}
\ba
\hspace*{-0.6cm}
\dd\sigma^{{\rm in}\;s}_{1L}
&=\!& \frac\alpha{2\pi} l_m 
\int\limits_{z_{1i}}^1 \dd z_1\, \big[P(z_1)\big]_+\, \frac{p_{ls}\, S_x^2}{p_l(z_1S-X)^2}
\nn
&\times&
\!\!\sigma_B(z_1S,z_1Q^2,x_s,z_s,p_{ts},\phi_{hs},\eta_{Ls},\phi_{\eta s}),
\\
\hspace*{-0.6cm}
\dd\sigma^{{\rm in}\;p}_{1L} 
&=& \frac\alpha{2\pi} l_m 
\int\limits_{z_{2i}}^1 \frac{\dd z_2}{z_2^2}\, \big[P(z_2)\big]_+\, \frac{p_{lp} S_x^2}{p_l(S-X/z_2)^2}
\nn
&\times&
\!\!\sigma_B(S,Q^2/z_2,x_p,z_p,p_{tp}, \phi_{hp},\eta_{Lp},\phi_{\eta p}) ,
\ea
\end{subequations}
are the respective $s$- and $p$-peak contributions.
The lower limits of integration are determined from the SIDIS pion threshold condition in Eq.~(\ref{thresh}),
\begin{subequations}
\label{z12m}
\ba
z_{1i} &=& 1 - \frac{p_X^2-M_{\rm th}^2}{S'}\, ,
\label{z1m}
\\
z_{2i} &=& \bigg( 1 + \frac{p_X^2-M_{\rm th}^2}{X'} \bigg)^{-1}. 
\label{z2m}
\ea
\end{subequations}
The splitting function $P(z)$ in Eqs.~(\ref{eq:dsigin1L}) is given by
\ba
P(z) = \frac{1+z^2}{1-z}, 
\label{spf}
\ea
and is evaluated according to the ``+ prescription'',
\ba
\int\limits_x^1 \dd z\, \big[ P(z )\big]_+\, f(z)
&=& \int\limits_x^1 \dd z\, P(z)\,\big[ f(z)-f(1) \big] 
\nn
&-& f(1) \int\limits_0^x \dd z\, P(z).
\label{eq:plus_def}
\ea  

For the representation of the shifted variables in the $s$- and $p$-peaks, it is useful to introduce the angles $\theta_{s,p}$ between the 3-momenta $\bm{q}$ and $\bm{q}_{s,p}$ in the target rest frame, respectively,
\begin{subequations}
\label{shift2}
\ba 
\cos\theta_s 
&=& \frac{\bm{q} \cdot \bm{q}_s}{|\bm{q}||\bm{q}_s|}
\nn
&=& \frac{(z_1S - X)\, S_x + 2 (z_1+1)\, M^2 Q^2}{\sqrt{\ly \lambda_{Ys}}},
\\
\sin\theta_s 
&=& \sqrt{1-\cos^2\theta_s}
\nn
&=& \frac{2 (1 - z_1)\, M \sqrt{Q^2(S X-M^2 Q^2)}}{\sqrt{\ly\lambda_{Ys}}},~~
\\
\cos\theta_p 
&=& \frac{\bm{q} \cdot \bm{q}_p}{|\bm{q}||\bm{q}_p|}
\nn
&=& \frac{(S - z_2^{-1} X)\, S_x + 2(1+z_2^{-1})\, M^2 Q^2}{\sqrt{\ly\lambda_{Yp}}},~~~~~
\\
\sin\theta_p 
&=& \sqrt{1-\cos^2\theta_p}
\nn
&=& \frac{2 (z_2^{-1} - 1)\, M \sqrt{Q^2(SX-M^2Q^2)}}{\sqrt{\ly\lambda_{Yp}}}.
\ea
\end{subequations}
\\
The components of the target polarization vector $\eta$ in the arguments of the Born cross section are determined in the $s$- and $p$-peaks when $q$ changes to
$q_s = z_1 k_1 - k_2$ and $q_p = k_1 - k_2/z_2$,
\begin{subequations}
\label{shift1}
\ba
\eta_{L\;s,p}&=&
\eta_L\cos\theta_{s,p}+\eta_t\cos\phi_\eta\sin\theta_{s,p}\, ,~~~~
\\
\sin\phi_{\eta \; s,p}&=&\frac{\eta_t}{\eta_{t\; s,p}}\sin\phi_\eta\, ,
\\
\cos\phi_{\eta \; s,p}&=&\frac{
\eta_t\cos\phi_\eta\cos\theta_{s,p}
-
\eta_L\sin\theta_{s,p}
}{\eta_{t\; s,p}}\, ,
\ea
\end{subequations}
with $\eta_{L\;s,p}^2 + \eta_{t\;s,p}^2 = 1$. 
We define $\ly=S_x^2+4M^2Q^2$, and the remaining variables with indices $s$ and $p$ in Eqs.~(\ref{shift2}) and (\ref{shift1}) are given by
\begin{subequations}
\label{shift3}
\ba
x_s &=& \frac{z_1 Q^2}{(z_1 S-X)}, \quad
z_s  =  \frac{z S_x}{(z_1 S-X)},\;
\\
x_p &=& \frac{Q^2}{(z_2 S-X)}, \quad
z_p  =  \frac{z S_x}{(S - X/z_2)},~~~~~~~~~~~
\\
p_{t\;s,p}^2 &=& \frac{z^2 S_x^2}{4M^2} - p_{l\;s,p}^2-m_h^2,
\\
p_{l\;s,p} &=& p_l \cos \theta_{s,p} - p_t \sin\theta_{s,p} \cos\phi_h,\;
\\
\lambda_{Ys} &=& (z_1 S-X)^2 + 4 z_1 M^2 Q^2,
\\
\lambda_{Yp} &=& \Big( S-\frac{X}{z_2} \Big)^2 + \frac{4}{z_2} M^2 Q^2,
\\
\cos\phi_{h\;s,p}
&=& \frac{p_t\cos\theta_{s,p} \cos\phi_h+p_l\sin\theta_{s,p} }{p_{t\;s,p}},
\\
\sin\phi_{h\;s,p}
&=&\frac{p_t\sin\phi_h}{p_{t\;s,p}}.
\ea
\end{subequations}

The contribution from vacuum polarization, $\delta_{\rm vac}^l$ in Eq.~(\ref{eq:siga_obs2}) [see Fig.~\ref{fgrc}~b)], does not involve splitting functions and must be added separately.
For electron loops, 
\ba
\delta_{\rm vac}^l(Q^2) = \frac23\, l_m\, .
\label{dvac}
\ea
while contributions from $\mu$ and $\tau$ leptons, and hadrons, are not included in the leading log formalism~\cite{Akushevich:2019mbz}.

The contribution from the ERT can be written in a closed form,
\ba
\sigma^{\rm ex}_{1L}=\sigma^{{\rm ex}\;s}_{1L}+\sigma^{{\rm ex}\;p}_{1L} ,
\label{sllex1}
\ea
where the $s$-peak and $p$-peak contributions are
\begin{subequations}
\label{sexll1}
\begin{align}
\sigma^{{\rm ex}\;s}_{1L} 
&= \frac{\alpha}{2\pi} l_m\,
\bigg( \frac{1+z_{1e}^2}{1-z_{1e}} \bigg)
\frac{p_{lse}}{p_l}\frac{S^2_x}{S'}
\left[ \frac{1}{z_{1e} S - X} + \frac{1}{2M^2} \right]
\nn
&\times
\bar\sigma^{\rm ex}_B(z_{1e}S,z_{1e}Q^2,x_{se},p_{tse},\phi_{hse},\eta^{\rm ex}_{Lse},\phi_{\eta se}) ,
\nn
&
\\
\sigma^{{\rm ex}\;p}_{1L} 
&= \frac{\alpha}{2\pi} l_m\,
\bigg( \frac{1+z_{2e}^2}{1-z_{2e}} \bigg)
\frac{p_{lpe}}{p_l}\frac{S^2_x}{X^\prime }
\left[ \frac 1{S-X/z_{2e}}+\frac 1{2M^2} \right]
\nn
&\times
\bar\sigma^{\rm ex}_B(S,Q^2/z_{2e},x_{pe},p_{tpe},\phi_{hpe},\eta^{\rm ex}_{Lpe},\phi_{\eta pe}) ,
\nn
&
\end{align}
\end{subequations}
and $\bar \sigma^{\rm ex}_B(S,Q^2,x,p_{t},\phi_h,\eta^{\rm ex}_L,\phi_\eta)$ is the four-fold Born exclusive cross section. 
The variables with indices ``$se$'' and ``$pe$'' are calculated from Eqs.~(\ref{shift2})--(\ref{shift3}) using the replacements:
\begin{itemize}
\item[(i)]  for $\sigma^{{\rm ex}\;s}_{1L}$: 
$z_1\to z_{1e}$ and \\ $z \to (M^2+m_h^2-m_u^2+z_{1e}S^\prime+V_2-X)/S_x$, 
\item[(ii)] for $\sigma^{{\rm ex}\;p}_{1L}$:
$z_2\to z_{2e}$ and \\ $z \to (M^2+m_h^2-m_u^2+X^\prime/z_{2e}+S-V_1)/S_x$.\\
\end{itemize}

\subsubsection{Electron structure functions \label{eq::RCs_lo3}}

\begin{figure}[t]
\centering
\scalebox{0.65}{\includegraphics{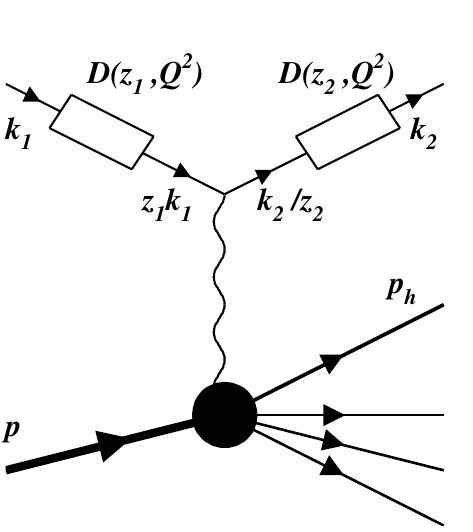}}
\caption{Representation of the SIDIS amplitude in the electron structure functions approach.}
\label{esf}
\end{figure}

Within the electron structure function method, illustrated in Fig.~\ref{esf}, the unpolarized SIDIS cross section is given by
\begin{align}
\sigma^{\rm in}_{hL} 
&= \frac{S_x^2}{p_l} 
\int\limits_{z_{1i}}^1 \dd z_1\, D(z_1,Q^2)
\int\limits_{{\hat z}_{2i}}^1 \frac{\dd z_2}{z^2_2}\,
D(z_2,Q^2)\, r^2 \bigg( \frac{z_1}{z_2} Q^2 \bigg)
\nn
& \hspace*{0.3cm} \times 
\frac{\hat p_l\, \sigma_{B}(z_1 S,z_1 Q^2/z_2, \hat x, \hat z, \hat p_t,\hat \phi_h, \hat \eta_L,\hat \phi_\eta)}{(z_1 S - X/z_2)^2},
\label{inll}
\end{align}
where $z_{1i}$ is defined in Eq.~(\ref{z1m}), and
\ba
\hat{z}_{2i} = \biggl[1+\frac{p_X^2-(1-z_1)S^\prime-M_{\rm th}^2}{X-V_2+z_1Q^2}\biggr]^{-1}.
\label{hz2}
\ea
The functions $D(z_{1,2},Q^2)$ here are referred to as the ``electron structure functions'' \cite{kuraev1,kuraev2,AAM2004}.
While these can more formally be referred to as electron distribution functions, for historical reasons we will refer to these as structure functions in the following.
The electron structure functions contain three terms, 
\ba
D\, =\, D^\gamma\, +\, D^{e^+e^-}_N\, +\, D^{e^+e^-}_S,
\label{esf3}
\ea
where $D^\gamma$ describes the photon radiation, and $D^{e^+e^-}_N$ and $D^{e^+e^-}_S$ represent electron pair production in the nonsinglet (by a single-photon mechanism) and singlet (by a double-photon mechanism) channels, respectively~\cite{kuraev1, kuraev2, AAM2004, ESFRAD}.
The explicit expressions of the components of $D(z_{1,2},Q^2)$ are given in Eqs.~(5)--(7) of Ref.~\cite{AAM2004}. 
The coefficient $r^2$ in the integrand of Eq.~(\ref{inll}) results from resummation of the vacuum polarization by leptons from Eq.~(\ref{dvac}), and can be expressed in terms of the QED coupling as
\ba
\!\!\!\!
r(Q^2)=\sum\limits_{i=0}^\infty\left(\frac{\alpha}{2\pi}\delta_{\rm vac}^l( Q^2)\right)^i= \biggl[1-\frac{\alpha}{2\pi}\delta_{\rm vac}^l( Q^2)\biggr]^{-1}.
\label{eq:vac_resum}
\ea

In analogy with Eqs.~(\ref{shift1}), the components of the target polarization vector $\eta$ in the arguments of the Born cross section are determined in the shifted variables, when the four-momentum transfer $q$ is replaced by $\hat q=z_1k_1-k_2/z_2$,
\begin{subequations}
\label{shift4}
\ba
\label{shift4_etaL}
\hat\eta_{L}
&=& \eta_L\cos\hat \theta+\eta_t\cos\phi_\eta\sin\hat \theta, 
\\
\label{shift4_cosphietaL}
\cos\hat\phi_\eta
&=& \frac{\eta_t\cos\phi_\eta \cos\hat\theta - \eta_L\sin\hat\theta}{\hat\eta_t},~~~~~~~~~
\\
\label{shift4_sinphietaL}
\sin\hat\phi_\eta
&=& \frac{\eta_t}{\hat\eta_t}\sin\phi_\eta,
\ea
\end{subequations}
where $\hat\eta_t^2 + \hat\eta_L^2 = 1$. 
Here, $\hat\theta$ is the angle between the 3-momentum vectors $\bm q$ and $\hat{\bm q}$ in the target rest frame, for which we have the relations,
\begin{subequations}
\label{shift5}
\begin{align}
\cos\hat\theta\, 
&= \frac{{\bm q}\cdot\hat{\bm q}}{|{\bm q}||\hat{\bm q}|}
\nn 
&= \frac{(z_1 S - X/z_2)\, S_x + 2 (z_1 + 1/z_2)\, M^2 Q^2}{\sqrt{\ly\hat\lambda_Y}},
\\
\sin\hat\theta\, 
&= \sqrt{1-\cos^2\hat\theta}
\nn
&= \frac{2(1/z_2-z_1)\, M \sqrt{Q^2(SX-M^2Q^2)}}{\sqrt{\ly\hat\lambda_Y}}.
\label{shift5}
\end{align}
\end{subequations}
The remaining variables with the carets ``$\hat{\phantom{x}}$'' are given by the relations
\begin{subequations}
\label{shift6}
\ba
\hat x &=& \frac{z_1Q^2}{(z_1z_2S-X)}, 
\hat z\,=\, \frac{z_2zS_x}{(z_2z_1S-X)},~~~~~~~~
\label{shift6_xz}
\\
\label{shift6_pt}
\hat p_t^2 &=& \frac{z^2S_x^2}{4M^2}-\hat p_l^2-m_h^2,
\\
\label{shift6_pl}
\hat p_l &=& p_l\cos \hat \theta-p_t\sin \hat \theta\cos \phi_h,
\\
\hat\lambda_Y &=& \Big( z_1 S - \frac{X}{z_2} \Big)^2 + \frac{4 z_1}{z_2} M^2 Q^2,
\\
\label{shift6_cosphihhat}
\cos\hat\phi_h
&=& \frac{p_t\cos\hat\theta \cos\phi_h + p_l\sin\hat\theta}{\hat p_t},
\\
\label{shift6_sinphihhat}
\sin\hat\phi_h 
&=& \frac{p_t\sin\phi_h}{\hat p_t}.
\ea
\end{subequations}
Finally, the contribution to the cross section from the ERT can be written as
\begin{align}
\sigma^{\rm ex}_{hL} 
&= \frac{S_x^2}{p_l}
\Bigg\{
\int\limits_{z_m}^1 \dd z_1\, 
\frac{D(z_1,Q^2)\, D(\hat z_{2},Q^2)}{X-V_2+z_1Q^2}\,
r^2 \bigg( \frac{z_1}{\hat z_{2}} Q^2 \bigg)
\nn
& \hspace*{1.2cm} \times
\hat p_{l2}
\biggl[ \frac 1{z_1 S-X/\hat z_2} + \frac 1{2M^2} \biggr]
\nn
& \hspace*{1.2cm} \times
\bar\sigma^{\rm ex}_B(z_1S, z_1 Q^2/z_2, \hat x_2, \hat p_{t2}, \hat\phi_{h2}, 
\hat\eta^{\rm ex}_{L2}, \hat\phi_{\eta 2})
\nn
& \hspace*{0.7cm} +
{\int\limits_{z_m}^1} \frac{\dd z_2}{z_2^2}\, 
\frac{D(\hat z_1,Q^2)\, D(z_2,Q^2)}{S-V_1-Q^2/z_2}\,
r^2 \bigg( \frac{\hat z_1}{z_2} Q^2 \bigg)
\nn
& \hspace*{1.2cm} \times
\hat p_{l1}
\biggl[ \frac 1{\hat z_1 S-X/z_2} + \frac 1{2M^2} \biggr]
\nn 
& \hspace*{1.2cm} \times
\bar \sigma^{\rm ex}_B(\hat z_1 S, \hat z_1 Q^2/z_2, \hat x_1, \hat p_{t1},\hat \phi_{h1}, 
\hat\eta^{\rm ex}_{L1},\hat\phi_{\eta 1})
\Bigg\} ,
\label{exll}
\end{align}
where
\begin{subequations}
\ba
\hat z_1 &=& 1-\frac{p_X^2 + (1-1/z_2) X' - m_u^2}{S - V_1 - Q^2/z_2},
\\
\hat z_2 &=& \biggl[ 1 + \frac{p_X^2 - (1-z_1)S' - m_u^2}{X -V_2 + z_1 Q^2} \biggr]^{-1},
\label{hz2e}
\ea
\end{subequations}
and the lower limits of integration is given by
\begin{align}
z_m &= \frac1{2(S-V_1)}
\bigg[
S_x + m^2_u - p_X^2 - 2V_-
\nn
+\ & \sqrt{(p_X^2 - S_x - m^2_u + 2 V_-)^2 + 4(S - V_1)(X - V_2)} 
\bigg],
\label{zmmm}
\end{align}
with $V_- = \frac12 (V_1-V_2)$.
The other shifted variables in Eq.~(\ref{exll}) are defined by Eqs.~(\ref{shift2})--(\ref{shift3}) with the transformations,
\begin{itemize}
\item[i)]  $z_2 \to \hat z_2$ and $z \to (M^2+m_h^2-m_u^2+z_1(S-V_1) \\ +(X-V_2)/\hat z_2)/S_x$ for the $z_1$ integral, 
\item[ii)] $z_1 \to \hat z_1$ and $z \to (M^2+m_h^2-m_u^2+\hat z_1(S-V_1) \\ +(X-V_2)/z_2)/S_x$ for the $z_2$ integral.
\end{itemize}

\subsection{Radiative effects in a factorized approach \label{eq::RCs_fac}}

As discussed in Sec.~\ref{eq::RCs_gen}, the factorized approach developed in Refs.~\cite{Liu:2020rvc, Liu:2021jfp} simultaneously treats QED and QCD effects on equal footing. 
In particular, instead of considering QED radiation as a correction to the Born process, in this approach the QED and QCD contributions to the SIDIS cross section are unified in a consistent factorization framework.

\subsubsection{Collinearly factorized QED contributions \label{eq::RCs_fac1}}

Since in the one-photon exchange approximation the virtual photon transverse momentum generated by leptonic radiation is much smaller than the typical intrinsic parton transverse momentum in hadrons~\cite{Liu:2021jfp}, the observable SIDIS cross section including QED contributions can be written in the factorized form~\cite{Liu:2020rvc, Liu:2021jfp},
\begin{align}
\label{RC_x_fac}
\frac{\dd^6\sigma^{\rm obs}}{\dd x\, \dd y\, \dd\phi\, \dd z\, \dd\phi_h\, \dd p_t^2}
&= \int\limits_{\xi_\text{min}}^1 \dd\xi\,  f_{e/e}(\xi,Q^2) 
\nn
&\hspace*{-3cm} \times 
\int\limits_{\zeta_\text{min}(\xi)}^1 \dd\zeta\, D_{e/e}(\zeta,Q^2)\,
\widehat{J}\,
\frac{\dd^6\hat{\sigma}\left(\alpha(Q^2)\right)}{\dd\hat{x}\, \dd\hat{y}\, \dd\hat{\phi}\, \dd\hat{z}\, \dd\hat{\phi}_h\, \dd\hat{p}_t^2}\, ,
\end{align}
where $\xi$ and $\zeta$ are the leptonic longitudinal momentum fractions in the lepton distribution and fragmentation functions, respectively, and
\begin{equation}
\widehat{J}\ \equiv\
\left\lVert \frac{\partial(\hat{x},\hat{y},\hat{\phi},\hat{z},\hat{\phi}_h,\hat{p}_{t}^2)}{\partial(x,y,\phi,z,\phi_h,p_{t}^2)}
\right\rVert 
\end{equation}
is the Jacobian.
The LDF $f_{e/e}(\xi,Q^2)$ and LFF $D_{e/e}(\zeta,Q^2)$ in Eq.~(\ref{RC_x_fac}) capture all infrared-sensitive contributions from collinear QED radiation associated with the initial- and final-state electrons, respectively, and have probability interpretations similar to those of hadronic PDFs and fragmentation functions.
In this section, the variables with carets denote functions of the leptonic longitudinal momentum fractions $\xi$ and $\zeta$.

In the infrared-safe cross section $\hat{\sigma}$, logarithmic QED contributions proportional to $\log^n(\mu_R/Q)$ are suppressed by setting the renormalization scale, $\mu_R$, to $Q^2$.
Assuming that other QED contributions are small, $\hat{\sigma}$ can be approximated by the Born cross section,
\begin{align}\label{eq:fac_born}
&\frac{\dd^6\hat{\sigma}\!\left(\alpha(Q^2)\right)}{\dd\hat{x}\,\dd\hat{y}\,\dd\hat{\phi}\,\dd\hat{z}\,\dd\hat{\phi}_h\,\dd\hat{p}_t^2} \nn
&\quad \approx \sigma_B\!\left(\hat{x},\hat{y},\widehat{Q}^2,\hat{\eta}_L,\hat{\eta}_t,\hat{\phi}_\eta,\hat{z},\hat{\phi}_h,\hat{p}_t^2;\alpha(Q^2)\right),
\end{align}
where $\sigma_B$ is given in Eq.~\eqref{born_xsection2} with the QED coupling $\alpha(Q^2)$ evaluated in the modified minimal subtraction ($\overline{\mathrm{MS}}$) scheme.

The leptonic longitudinal momentum fractions $\xi$ and $\zeta$ are defined as
\begin{equation}
\hat{k}_1 = \xi\, k_1, \qquad 
\hat{k}_2 = \frac{1}{\zeta}\, k_2,
\end{equation}
where $\hat{k}_1$ and $\hat{k}_2$ are the fractional initial- and final-state lepton momenta that enter the hard process described by $\hat{\sigma}$.
Within the factorized framework, the true virtual photon that probes the hadron has momentum
\begin{equation}
\hat{q} = \hat{k}_1 - \hat{k}_2.
\end{equation}
The cross section $\hat\sigma$ is evaluated with shifted variables defined in the true photon-nucleon frame, rather than in the experimentally defined photon-nucleon frame.

The formulas relating the variables in the two frames are derived in the appendix~C in Ref.~\cite{Liu:2021jfp}. Here we list the ones relevant for the present discussion, beginning with the Lorentz invariants,
\begin{subequations}
\label{eq:fac_shifted}
\begin{align}
\label{eq:fac_shifted_xz}
&\hat{x}\ =\ \frac{\xi xy}{\xi\zeta-(1-y)}, \qquad 
\hat{z}\, =\, \frac{\zeta zy}{\xi\zeta-(1-y)}, 
\\
\label{eq:fac_shifted_Qygamma}
&\widehat{Q}^2\, =\, \frac{\xi}{\zeta}\, Q^2, \quad 
\hat{y} = 1-\frac{1-y}{\xi\zeta}, \quad 
\hat{\gamma} = \frac{2M\hat{x}}{\widehat{Q}}, 
\\
\label{eq:fac_shifted_pt}
&\hat{p}_t^2\, =\, \frac{\hat{z}^2 \widehat{Q}^4 + 2(\hat{q}\cdot p_h)\, \hat{z}\, \widehat{Q}^2 - \hat{\gamma}^2(\hat{q}\cdot p_h)^2}{(1+\hat{\gamma}^2)\, \widehat{Q}^2}
- m_h^2, 
\end{align}
\end{subequations}
and azimuthal angles
\begin{subequations}
\begin{align}
\label{eq:fac_shifted_cosphih}
\cos\hat{\phi}_h &= \frac{1}{\hat{k}_{1t}\, \hat{p}_{t}} 
\bigg[ - \xi\, k_1\cdot p_h 
\nonumber\\
& \hspace*{0.3cm}
+ \frac{\big(1+\frac12\hat{y}\hat{\gamma}^2\big)(\hat{q}\cdot p_h) 
        + \big(1-\frac12\hat{y}\big)\, \hat{z}\, \widehat{Q}^2}
       {\hat{y}(1+\hat{\gamma}^2)} 
\bigg], 
\\
\label{eq:fac_shifted_sinphih}
\sin\hat{\phi}_h &= \frac{p_t}{\hat{p}_t} \sin\phi_h, 
\\
\label{eq:fac_shifted_cosphieta}
\cos{\hat{\phi}_\eta} &= \frac{1}{\hat{k}_{1t}\hat{\eta}_{t}} 
\bigg[ 
- \xi\, k_1\cdot\eta + \frac{1+\frac12\hat{y}\, \hat{\gamma}^2}{\hat{y}\,(1+\hat{\gamma}^2)}\, \hat{q}\cdot\eta 
\bigg], 
\\
\label{eq:fac_shifted_sinphieta}
\sin{\hat{\phi}_\eta} &= \frac{\eta_t}{\hat{\eta}_t} \sin\phi_\eta,
\end{align}
\end{subequations}
with the spin projections given by
\begin{align}
\label{eq:fac_shifted_etaL}
\hat{\eta}_L &= \frac{\hat{\gamma}(\hat{q}\cdot\eta)}{\sqrt{1+\hat{\gamma}^2}~\widehat{Q}}, 
\qquad 
\hat{\eta}_t^2 = 1 - \hat{\eta}_L^2.
\end{align}
The scalar products in Eqs.~(\ref{eq:fac_shifted})--(\ref{eq:fac_shifted_etaL}) are given by
\begin{subequations}
{\allowdisplaybreaks
\begin{align}
&\hat{q}\cdot p_h 
= \Big(\xi-\frac{1}{\zeta}\Big) k_1\cdot p_h + \frac{1}{\zeta} q\cdot p_h, 
\\
&q\cdot p_h = \frac{1}{\gamma^2} \bigg( z Q^2 
\nn
&\qquad 
-\sqrt{(1+\gamma^2)\, Q^2\big(z^2 Q^2 - \gamma^2 (m_h^2+p_t^2)\big)} 
\bigg), 
\\
&k_1\cdot p_h 
= - k_{1t}\, p_t \cos\phi_h 
\nn
&\qquad+ \frac{\big(1+\frac12 y\gamma^2\big)(q\cdot p_h) + \big(1-\frac12 y\big) z\, Q^2}{y\, (1+\gamma^2)}, 
\\
&\hat{q}\cdot\eta = \Big( \xi-\frac{1}{\zeta} \Big) k_1\cdot\eta + \frac{1}{\zeta} q\cdot\eta, 
\\
&q\cdot\eta = \frac{1}{\gamma} \sqrt{1+\gamma^2}~Q\, \eta_L, 
\\
&k_1\cdot\eta 
= - k_{1t}\, \eta_t \cos\phi_\eta + \frac{1+\frac12 y\gamma^2}{y\, (1+\gamma^2)}\, q\cdot\eta, 
\\
&k_{1t}^2 
= \frac{1-y-\frac14 y^2\gamma^2}{y^2 (1+\gamma^2)} Q^2,
\\
&\hat{k}_{1t}^2
= \frac{1-\hat{y}-\frac14 \hat{y}^2 \hat{\gamma}^2}{\hat{y}^2 (1+\hat{\gamma}^2)} \widehat{Q}^2.
\end{align}}%
\end{subequations}%
For the polarization vector $\eta$, one often uses the approximation $\phi_\eta \approx \phi$, obtained in the limit $M/Q \to 0$.
In this limit, one also has
\begin{align}
\hat{\eta}_L \approx \eta_L, \qquad 
\hat{\eta}_T \approx \eta_T, \qquad 
\hat{\phi}\approx\hat{\phi}_\eta \approx \phi_\eta.
\end{align}
The Jacobian in Eq.~\eqref{RC_x_fac} can be written as
\begin{equation}
\widehat{J}\ =\ \frac{y^2}{\big( \xi\zeta-(1-y) \big)^2}
\sqrt{\frac{1-\hat{\gamma}^2(m_h^2+\hat{p}_t^2)/(\hat{z}^2\widehat{Q}^2)}{1-\gamma^2(m_h^2+p_t^2)/(z^2Q^2)}}.
\label{eq:jac_factor}
\end{equation}

The lower limits of the momentum fractions $\xi$ and $\zeta$ in Eq.~\eqref{RC_x_fac} are derived from the SIDIS pion threshold condition $(\hat{q}+p-p_h)^2>M_{\rm th}^2$ [see Eq.~\eqref{thresh}], and are given by%
\footnote{In Ref.~\cite{Liu:2021jfp} the lower limits $\xi_\text{min}$ and $\zeta_\text{min}(\xi)$ were derived from the kinematical restrictions $\hat{x}_B < 1$ and $\hat{z}_h < 1$, which are necessary (but not sufficient) conditions for the SIDIS pion threshold condition $(\hat{q}+p-p_h)^2 > M_{\rm th}^2$.}
\begin{align}
\label{eq:low_limits_factor1}
&\xi_\text{min}
= \frac{1 - y + zy + xy\, (\Delta - 2\, k_2 \cdot p_h)/Q^2}
       {1 - xy\, (1 + 2 k_1 \cdot p_h/Q^2)},
\\
\label{eq:low_limits_factor2}
&\zeta_\text{min}(\xi)
= \frac{1-y+\xi xy - 2\, xy\, k_2 \cdot p_h/Q^2}
  {\xi - zy - xy\, (\Delta + 2\, \xi\, k_1\cdot p_h)/Q^2},
\end{align}
where $\Delta \equiv M_{\rm th}^2-(M^2+m_h^2)$ and $k_2\cdot p_h = k_1\cdot p_h - q\cdot p_h$.
Below the SIDIS pion threshold there are contributions from the exclusive channel, which are not yet included in the factorized framework (but are implemented in the exact and traditional leading-order framework discussed in Sec.~\ref{eq::RCs_lo}).

Unlike the nonperturbative hadronic PDFs and fragmentation functions, the LDFs and LFFs can be calculated perturbatively, for the case of radiated photons splitting into lepton--antilepton pairs.
In the case of photons splitting into quark--antiquark pairs, or at higher orders, long-distance QCD effects introduce nonperturbative contributions to the LDFs and LFFs~\cite{Cammarota:2025jyr,Qiu:2026fed}.
Nevertheless, due to their universality, it is expected that these functions can be extracted from experiment.

In the factorized framework, collinear QED radiation gives rise to large logarithms of $\mu/m$ at each order in $\alpha$. 
To leading-log accuracy, these are resummed into the LDF and LFF in Eq.~\eqref{RC_x_fac}, to all orders, by solving the QED analogue of the DGLAP evolution equations~\cite{Dsp, GLsp, Lipatov:1974qm, APsp},
\begin{subequations}
\label{eq:DGLAP}
\begin{align}
\hspace*{-0.2cm}
\label{eq:DGLAP_LDF}
\frac{\dd f_{i/e}(\xi,Q^2)}{\dd\log Q^2} &= \frac{\alpha(Q^2)}{2\pi} \sum_j P_{ij}(\xi) \otimes_\xi f_{j/e}(\xi,Q^2), 
\\
\hspace*{-0.2cm}
\label{eq:DGLAP_LFF}
\frac{\dd D_{e/i}(\zeta,Q^2)}{\dd\log Q^2} &= \frac{\alpha(Q^2)}{2\pi} \sum_j D_{e/j}(\zeta,Q^2) \otimes_\zeta P_{ji}(\zeta),
\end{align}
\end{subequations}
where $i,j=e,\bar{e},\gamma$, and the symbol $\otimes_x$ denotes the Mellin convolution,
\begin{align}
f(x) \otimes_x g(x) \equiv \int\limits_x^1\frac{\dd z}{z}\, f(z)\, g\Big(\frac{x}{z}\Big).
\end{align}

Note that the LDF and the LFF peak as $\xi \to 1$ and $\zeta \to 1$, respectively.
To obtain accurate numerical results for the integrals in Eq.~\eqref{RC_x_fac}, one solves Eqs.~(\ref{eq:DGLAP}) in Mellin space and using a subtraction trick~\cite{Liu:2021jfp} to obtain,
\begin{align}\label{eq:subtrick1}
&\frac{\dd^6\sigma^{\rm obs}}{\dd x\, \dd y\, \dd\phi\, \dd z\, \dd\phi_h\, \dd p_t^2} 
\nn
&\quad= \int\limits_{\xi_\text{min}}^1 \dd\xi\, f_{e/e}(\xi,Q^2) \int\limits_{\zeta_\text{min}(\xi)}^1 \dd\zeta\, D_{e/e}(\zeta,Q^2)\, \mathcal{H}(\xi,\zeta) 
\nn
&\quad= \int\limits_{\xi_\text{min}}^1 \dd\xi\, f_{e/e}(\xi,Q^2) \big[\mathcal{G}(\xi)-\mathcal{G}(1)\big] 
\nn
&\qquad+ \mathcal{G}(1) \int\limits_{\xi_\text{min}}^1 \dd\xi\, f_{e/e}(\xi,Q^2),
\end{align}
where
\begin{align}\label{eq:subtrick2}
\mathcal{G}(\xi) 
&\equiv \int\limits_{\zeta_\text{min}(\xi)}^1 
\dd\zeta\, D_{e/e}(\zeta,Q^2)\,
\big[ \mathcal{H}(\xi,\zeta)-\mathcal{H}(\xi,1) \big] 
\nn
&\quad 
+ \mathcal{H}(\xi,1) \int\limits_{\zeta_\text{min}(\xi)}^1 \dd\zeta\, D_{e/e}(\zeta,Q^2),
\end{align}
and $\mathcal{H}(\xi,\zeta)$ contains the Jacobian and the cross section $\hat{\sigma}$ in Eq.~\eqref{RC_x_fac}.
The final-line integrals in Eqs.~\eqref{eq:subtrick1} and \eqref{eq:subtrick2} are evaluated as
\begin{align}\label{eq:LDFintegral}
&\int\limits_{x_\text{min}}^1 \dd x\, F(x) 
= x_\text{min}\, \mathcal{M}^{-1}\!\biggl[\frac{F^N}{N-1}\biggr](x_\text{min}),
\end{align}
where $F^N$ denotes the $N$-th Mellin moment of $F(x)$,
\begin{align}\label{eq:MellinTransform}
F^N = \mathcal{M}[F(x)](N) = \int\limits_0^1 \dd x \, x^{N-1} F(x),
\end{align}
and $\mathcal{M}^{-1}$ represents the inverse Mellin transform,
\begin{align}\label{eq:InvMellinTransform}
F(x) = \mathcal{M}^{-1}\big[F^N\big](x) = \frac{1}{2\pi i} \int \dd N\, x^{-N} F^N.
\end{align}
The Mellin moments of the LDFs and LFFs are calculated analytically, while the inverse Mellin transform is calculated numerically using Eq.~\eqref{eq:InvMellinTransformNum}.

Solving for the running electromagnetic coupling in the $\overline{\text{MS}}$ scheme, we find, at leading-log accuracy,
\begin{align}\label{eq:alphaMSbar}
\alpha(Q^2) = \alpha(m^2) \left(1+\alpha(m^2)\, \beta_0 \log\frac{Q^2}{m^2} \right)^{-1},
\end{align}
where $\alpha(m^2)=1/137$ and $\beta_0=-1/(3\pi)$.
Defining the nonsinglet ($f^N_n$, $D^N_n$) and singlet ($f^N_s$, $D^N_s$) functions as
\begin{subequations}
\begin{align}
&
f_n^N = f_{e/e}^N - f_{\bar{e}/e}^N, \qquad 
D_n^N = D_{e/e}^N - D_{e/\bar{e}}^N, 
\\
&
f_s^N = f_{e/e}^N + f_{\bar{e}/e}^N, \qquad 
D_s^N = D_{e/e}^N + D_{e/\bar{e}}^N,
\end{align}
\end{subequations}
the solutions to Eqs.~\eqref{eq:DGLAP} for the LDFs and LFFs in Mellin space are given by (see Appendix~\hyperref[sec:appA]{A}),
\begin{subequations}
\label{eq:LDF_solution}
\begin{align}
\label{eq:LDFv_solution}
f_n^N(Q^2) 
&= r(Q^2)^{-P_{ee}^N/(2\pi\beta_0)}\, f_n^N(m^2), 
\\
\label{eq:LDFs_solution}
\begin{pmatrix} f_s^N(Q^2) \\ f_{\gamma/e}^N(Q^2) \end{pmatrix} 
&= r(Q^2)^{-\mathbf{P}_s^N/(2\pi\beta_0)} \begin{pmatrix} f_s^N(m^2) 
\\ f_{\gamma/e}^N(m^2) \end{pmatrix}, 
\\
\label{eq:LFFv_solution}
D_n^N(Q^2) 
&= r(Q^2)^{-P_{ee}^N/(2\pi\beta_0)}\, D_n^N(m^2), 
\\
\label{eq:LFFs_solution}
\begin{pmatrix} D_s^N(Q^2) \\ D_{e/\gamma}^N(Q^2) \end{pmatrix} 
&= r(Q^2)^{-\mathbb{P}_s^N/(2\pi\beta_0)} \begin{pmatrix} D_s^N(m^2) \\ D_{e/\gamma}^N(m^2) \end{pmatrix},
\end{align}
\end{subequations}
where $r(Q^2) \equiv \alpha(Q^2)/\alpha(m^2)$, and
\begin{align}
\label{eq:splitfns}
\mathbf{P}_s^N \equiv \begin{pmatrix} P_{ee}^N & 2P_{e\gamma}^N \\ P_{\gamma e}^N & P_{\gamma\gamma}^N \end{pmatrix}, \quad
\mathbb{P}_s^N \equiv \begin{pmatrix} P_{ee}^N & 2P_{\gamma e}^N \\ P_{e\gamma}^N & P_{\gamma\gamma}^N \end{pmatrix}.
\end{align}
The splitting kernels relevant for the solutions in Eqs.~\eqref{eq:LDF_solution}--\eqref{eq:splitfns} are given by
\begin{subequations}
\label{eq:split_factor}
\begin{align}
\label{eq:split_factor_ee}
&P_{ee}(x) = \left[\frac{1+x^2}{1-x}\right]_+, \\
\label{eq:split_factor_egamma}
&P_{e\gamma}(x) = x^2+(1-x)^2, \\
\label{eq:split_factor_gammae}
&P_{\gamma e}(x) = \frac{1+(1-x)^2}{x}, \\
\label{eq:split_factor_gammagamma}
&P_{\gamma\gamma}(x) = - \frac{2}{3} \delta(x-1),
\end{align}
\end{subequations}
where the subscript $+$ in Eq.~(\ref{eq:split_factor_ee}) denotes the plus prescription, as in Eq.~(\ref{eq:plus_def}).

The explicit expressions for the matrix exponential are given in Eqs.~\eqref{eq:exp_spacelikeP} and \eqref{eq:exp_timelikeP} of Appendix~A.
In terms of the Mellin moments, the LDF $f_{e/e}(\xi,Q^2)$ and LFF $D_{e/e}(\zeta,Q^2)$ in Eq.~\eqref{RC_x_fac} can be written as
\begin{subequations}
\label{eq:LFee_solution}
\begin{align}
\label{eq:LDFee_solution}
f_{e/e}(\xi,Q^2) &= \frac12 \mathcal{M}^{-1}
\big[ f_n^N(Q^2) + f_s^N(Q^2) \big]\!(\xi), 
\\
\label{eq:LFFee_solution}
D_{e/e}(\zeta,Q^2) &= \frac12 \mathcal{M}^{-1}
\big[ D_n^N(Q^2) + D_s^N(Q^2) \big]\!(\zeta),
\end{align}
\end{subequations}
where $\mathcal{M}^{-1}$ denotes the inverse Mellin transform defined in Eq.~\eqref{eq:InvMellinTransform}.

In Eq.~\eqref{eq:LDF_solution} the LDFs and LFFs at the initial scale are obtained by setting $Q^2 = m^2$ in the $\overline{\text{MS}}$ results at order 
$\mathcal{O}(\alpha)$~\cite{Frixione:2019lga},
\begin{subequations}
\label{eq:NLOLDF}
\begin{align}
\label{eq:NLOLDFee}
&f_{e/e}(x,m^2) 
 = D_{e/e}(x,m^2) 
 = \delta(x-1) 
\nn
& \hspace*{1.7cm}
+ \frac{\alpha(m^2)}{2\pi} 
  \left[ \frac{1+x^2}{1-x} \Big( \log\frac{1}{(1-x)^2} - 1 \Big) \right]_+, 
\\
\label{eq:NLOLDFgamma}
&f_{\gamma/e}(x,m^2) 
= \frac{\alpha(m^2)}{2\pi} 
  \left[\frac{1+(1-x)^2}{x}\right]
\Big(\log \frac{1}{x^2} - 1\Big), 
\\
\label{eq:NLOLFFgamma}
&D_{e/\gamma}(x,m^2) 
 = 0, 
\\
\label{eq:NLOLDFothers}
&f_{\bar{e}/e}(x,m^2)
= D_{e/\bar{e}}(x,m^2) = 0.
\end{align}
\end{subequations}%
The Mellin transforms \eqref{eq:MellinTransform} of these functions are presented in Eqs.~\eqref{eq:NLOLDFee_N}.

\subsubsection{First-order QED contributions}

For comparison with the leading-log result obtained in the traditional approach, as presented in Sec.~\ref{eq::RCs_lo2}, we derive the first-order leading-log (LL) formula for QED radiative effects in the factorized framework by expanding Eq.~\eqref{RC_x_fac} through order $\alpha\log(Q^2/m^2)$,
\begin{align}
&\frac{\dd^6\sigma^{\rm obs}_{\mathcal{O}(\alpha),\rm LL}}{\dd x\, \dd y\, \dd\phi\, \dd z\, \dd\phi_h\, \dd p_t^2} 
= \sigma_B + \frac{\alpha(m^2)}{2\pi}\log\frac{Q^2}{m^2} 
\Bigg[ \
\frac43\ \sigma_B
\nn
& \hspace*{0cm}
+ \!\!\int\limits_{\xi_\text{min}}^1 \!\!\!
  \dd\xi \bigg[ \frac{1+\xi^2}{1-\xi} \bigg]_+ 
  \!\widehat{J}\ \hat\sigma_B\Big|_{\zeta=1} 
+ \!\!\!\int\limits_{\zeta_\text{min}(1)}^1 \!\!\!\!\!
  \dd\zeta \bigg[ \frac{1+\zeta^2}{1-\zeta} \bigg]_+ 
  \!\widehat{J}\ \hat\sigma_B\Big|_{\xi=1} 
\Bigg],
\label{eq:xs_alpha_factor}
\end{align}
where 
$\sigma_B \equiv \sigma_B\big(x,y,Q,\eta_L,\eta_t,\phi_\eta,z,\phi_h,p_t^2;\alpha(m^2)\big)$
and
$\hat\sigma_B \equiv
\sigma_B\big(\hat{x},\hat{y},\widehat{Q}^2,\hat{\eta}_L,\hat{\eta}_t,\hat{\phi}_\eta,\hat{z},\hat{\phi}_h,\hat{p}_t^2;\alpha(m^2)\big)$,
and we have used Eq.~\eqref{eq:alphaMSbar} for the expansion of $\alpha(Q^2)$.
Including also the first-order non-log (NL) contributions from the LDF $f_{e/e}$ and LFF $D_{e/e}$ in Eq.~\eqref{eq:NLOLDFee}, we can write
\begin{align}
&\frac{\dd^6\sigma^{\rm obs}_{\mathcal{O}(\alpha),\rm LL+NL}}{\dd x\, \dd y\, \dd\phi\, \dd z\, \dd\phi_h\, \dd p_t^2} 
= \sigma_B + \frac{\alpha(m^2)}{2\pi} 
\Bigg[ \
\frac43 \log\frac{Q^2}{m^2}\ \sigma_B
\nn
&+ \int\limits_{\xi_\text{min}}^1 \dd\xi 
\bigg[ 
  \frac{1+\xi^2}{1-\xi} 
  \Big( \log\frac{Q^2}{(1-\xi)^2 m^2} - 1 \Big) 
\bigg]_+
\widehat{J}\ \hat\sigma_B\Big|_{\zeta=1}
\nn
&+ \!\!\!\int\limits_{\zeta_\text{min}(1)}^1 \!\!\!\dd\zeta 
\bigg[
  \frac{1+\zeta^2}{1-\zeta} 
  \Big( \log\frac{Q^2}{(1-\zeta)^2 m^2} - 1 \Big)
\bigg]_+ 
\widehat{J}\, \hat\sigma_B\Big|_{\xi=1} 
\Bigg].
\label{eq:xs_alpha_factor2}
\end{align}

\subsubsection{Other QED contributions}

In addition to the contributions described above, there are several other QED corrections to the SIDIS Born cross section in the factorized framework that enter in the one-photon approximation, but are
not incorporated in the present analysis. These include:
\begin{enumerate}[label=(\arabic*)]
\item Vacuum polarization effects of the exchanged virtual photon.
\item Higher-order and non-log QED contributions to the leptonic part. As shown in Sec.~3.3.2 of Ref.~\cite{Liu:2021jfp}, the leptonic tensor can be calculated with QED collinear (TMD) factorization for large (small) virtual photon transverse momentum in the lepton back-to-back frame, where large logs from soft and collinear radiations can be resummed to all orders.
\item QED contributions to the hadronic part. These include QED virtual corrections to the quark vertex, emissions of real hard photons from quarks, and collinear QED radiations from quarks.  These can be included by directly extending the QCD factorization theorems for the SIDIS hadronic tensor to include QED.
\end{enumerate}

\section{Results \label{eq::RCs_res}}

In this section we compare the results from the two general approaches discussed in Secs.~\ref{eq::RCs_lo} and \ref{eq::RCs_fac} above, both for the  analytical formulas and numerical evaluations for several SIDIS observables.

\subsection{Analytical comparison of the key formulas 
\label{eq::RCs_comp_form}} 

Comparing the theoretical formulas for the SIDIS cross sections in the exact + leading-order (traditional) approach and the factorized approach, we find general agreement between the results, when the electron mass is set to zero in the evaluation of the infrared-safe Born cross section in the factorized framework.
In particular, we consider the following ingredients:

\medskip
\paragraph{Cross sections.}
The cross section formula in Eq.~(\ref{inll}) is structurally similar to that in Eqs.~(\ref{RC_x_fac})--(\ref{eq:fac_born}).
The first-order leading-log cross section formula in Eqs.~(\ref{xs_contin})--(\ref{eq:dsigin1L}) is also structurally similar to that in Eq.~(\ref{eq:xs_alpha_factor}).

\medskip
\paragraph{Vacuum polarization and running QED coupling.}
In Eqs.~(\ref{RC_x_fac})--(\ref{eq:fac_born}), the running QED coupling is calculated using Eq.~\eqref{eq:alphaMSbar}.
On the other hand, Eq.~(\ref{inll}) uses a QED coupling at the initial scale and multiplies the cross section by a factor $r^2(Q^2)$ in Eq.~\eqref{eq:vac_resum} that encodes the vacuum polarization effects.
The two treatments give identical results at the cross section level.

\medskip
\paragraph{Shifted variables.}
The Born cross section $\sigma_B$ in Eqs.~\eqref{inll} and \eqref{eq:fac_born} is evaluated at the shifted variables $\hat{x}$, $\hat{y}$, $\widehat{Q}$, $\hat{z}$, $\hat{\eta}_L$, $\hat{\phi}_\eta$, $\hat{p}_t$,  and $\hat{\phi}_h$. We verify the equivalence of the corresponding expressions in the two frameworks by recasting all relevant quantities in terms of their unshifted (Born) counterparts: \\
\hspace*{0.2cm} $\bullet$
$\hat{x}$ and $\hat{z}$ in Eq.~\eqref{shift6_xz} are identical to those in Eq.~\eqref{eq:fac_shifted_xz}. \\
\hspace*{0.2cm} $\bullet$
$\widehat{Q}$ in Eq.~\eqref{eq:fac_shifted_Qygamma} is identical to the second argument of $\sigma_B$ in Eq.~\eqref{inll}. \\
\hspace*{0.2cm} $\bullet$
The formulas for $\hat{y}$ are identical. The respective formula in the traditional framework is not presented here, but can be deduced from the shifted $S$ variable (first argument of $\sigma_B$ in Eq.~\eqref{inll}), and can be found in Eq.~(3) of Ref.~\cite{AAM2004}. It matches the expression in Eq.~\eqref{eq:fac_shifted_Qygamma} in the factorized framework. \\
\hspace*{0.2cm} $\bullet$
$\hat{p}_t$, $\cos\hat{\phi}_h$, and $\sin\hat{\phi}_h$ in Eqs.~\eqref{shift6_pt}, \eqref{shift6_cosphihhat}, and \eqref{shift6_sinphihhat} are identical to those in Eqs.~\eqref{eq:fac_shifted_pt}, \eqref{eq:fac_shifted_cosphieta}, and \eqref{eq:fac_shifted_sinphieta}, respectively. \\
\hspace*{0.2cm} $\bullet$
$\hat{\eta}_L$, $\cos\hat{\phi}_\eta$, and $\sin\hat{\phi}_\eta$ in Eqs.~\eqref{shift4_etaL}, \eqref{shift4_cosphietaL}, and \eqref{shift4_sinphietaL} are identical to those in 
Eqs.~\eqref{eq:fac_shifted_etaL}, \eqref{eq:fac_shifted_cosphieta}, and \eqref{eq:fac_shifted_sinphieta}, respectively.

\medskip
\paragraph{Jacobian transformation.} The Jacobian in Eq.~(\ref{eq:jac_factor}) (used in Eqs.~(\ref{RC_x_fac}) and (\ref{eq:xs_alpha_factor})) is identical to that in Eqs.~(\ref{eq:dsigin1L}) 
and (\ref{inll}).

\medskip
\paragraph{Lower limits of integration.} The lower limits of integration in Eqs.~(\ref{eq:low_limits_factor1}) and (\ref{eq:low_limits_factor2}) are identical to those in Eqs.~(\ref{z12m}) and (\ref{hz2}) 
for zero electron mass, $m=0$.

\medskip
\paragraph{Splitting functions.} The splitting functions in Eq.~(\ref{spf}) and Eq.~(\ref{eq:split_factor_ee}) (and those used in Eq.~(\ref{eq:xs_alpha_factor})) are equivalent.

\subsection{Numerical comparison of SIDIS observables}
\label{eq::RCs_comp_obs}

Having established the correspondence between the analytical results from the traditional and factorized frameworks, we now compare the numerical results for RCs to SIDIS observables for $\pi^+$ electroproduction off the proton. We focus in particular on the unpolarized SIDIS cross section, and the Collins and Sivers transverse SSAs. 

The unpolarized SIDIS Born cross section is computed from Eq.~({\ref{born_xsection2}), setting the terms proportional to $\lambda_e$, $\eta_{L}$, and $\eta_{t}$ to zero, while the observed (radiation-included)
cross sections in the traditional and factorized approaches are computed using the formulas in Secs.~\ref{eq::RCs_lo} and \ref{eq::RCs_fac}, respectively. To quantify the RCs to the SIDIS cross section, it will 
be convenient to define the RC factor $\delta_{\rm RC}$ as a ratio of the observed cross section to the Born cross section,
\begin{equation}
\delta_{\rm RC} = \frac{\sigma^{\rm obs}}{\sigma_B}.
\label{eq:delta_RC_factor}
\end{equation}

For quantification of the RCs to the Sivers $A_{UT}^{\sin(\phi_h-\phi_\eta )}$ and Collins $A_{UT}^{\sin(\phi_h+\phi_\eta)}$ SSAs, we consider asymmetries defined by the ratios,
\begin{subequations}
\label{eq:AUT}
\ba \!\!\!\!\!\!\!\!
A_{UT}^{\sin(\phi_h-\phi_\eta)} =
\frac{2\displaystyle\int\limits_0^{2\pi}\dd\phi_h\int\limits_0^{2\pi}\dd\phi_\eta\,\sin(\phi_h-\phi_\eta)\,\sigma}
{\displaystyle\int\limits_0^{2\pi}\dd\phi_h\int\limits_0^{2\pi}\dd\phi_\eta\,\sigma}\, ,
\label{eq:AUT_Sivers}
\ea
\ba \!\!\!\!\!\!\!\!
A_{UT}^{\sin(\phi_h+\phi_\eta)} =
\frac{2\displaystyle\int\limits_0^{2\pi} \dd\phi_h
\int\limits_0^{2\pi} \dd\phi_\eta\,
\sin(\phi_h+\phi_\eta)\, \sigma}
{\displaystyle \epsilon \int\limits_0^{2\pi} \dd\phi_h\int\limits_0^{2\pi}\dd\phi_\eta\,\sigma}\, ,
\label{eq:AUT_Collins}
\ea
\end{subequations}
where $\sigma$ here refers to either the Born cross section, $\sigma_{B}$, or the observed (radiation included) cross section, $\sigma^{\rm obs}$.
At the Born level, the SSAs can be expressed as ratios of the semi-inclusive structure functions $F_{UT}^{\sin(\phi_h+\phi_\eta)}$ and $F_{UT,T}^{\sin(\phi_h-\phi_\eta)}$ to the unpolarized structure function $F_{UU,T}$, respectively,
\begin{subequations}
\label{scasym}
\ba
A_{UT}^{\sin(\phi_h-\phi_\eta)} &=& \frac{F_{UT,T}^{\sin(\phi_h-\phi_\eta)}}{F_{UU,T}},
\label{scasym1}
\\
A_{UT}^{\sin(\phi_h+\phi_\eta)} &=& \frac{F_{UT}^{\sin(\phi_h+\phi_\eta)}}{F_{UU,T}}.
\label{scasym2}
\ea
\end{subequations}

In all computations, only five structure functions are kept nonzero: the three unpolarized functions $F_{UU,T}$, $F_{UU}^{\cos \phi_h}$, and $F_{UU}^{\cos 2\phi_h}$, and the functions determining the Sivers and Collins SSAs, $F_{UT,T}^{\sin(\phi_h-\phi_\eta)}$ and $F_{UT}^{\sin(\phi_h+\phi_\eta)}$, respectively.
Each of these functions is computed according to the leading twist (Wandzura–Wilczek) approximation~\cite{Bastami:2018xqd, Proc_WW}, using the MSTW2008~\cite{Martin:2009iq} parametrization for the unpolarized proton PDFs, the NNPDFpol1.1~\cite{Nocera:2014gqa} fit for the proton helicity PDFs, and DSS07~\cite{deFlorian:2007aj} for the parton $\to$ pion fragmentation functions. 
The contribution of the ERT is calculated in the traditional approach using the MAID2007 parametrization~\cite{MAID2007helic} (see Ref.~\cite{Tra_RC_exc} for details).}
For each approach different codes have been used to compute the RCs: the traditional method uses the HAPRAD~2.0 code~\cite{RC_desk}, while the results from the factorized method were obtained using the Julia packages SIDISXSec~\cite{SIDISXSec} and QEDFactorization~\cite{QEDFactorization}.

\begin{figure*}[p]\centering
\scalebox{0.85}{\includegraphics[width=0.60\textwidth, height=1.05\textheight]{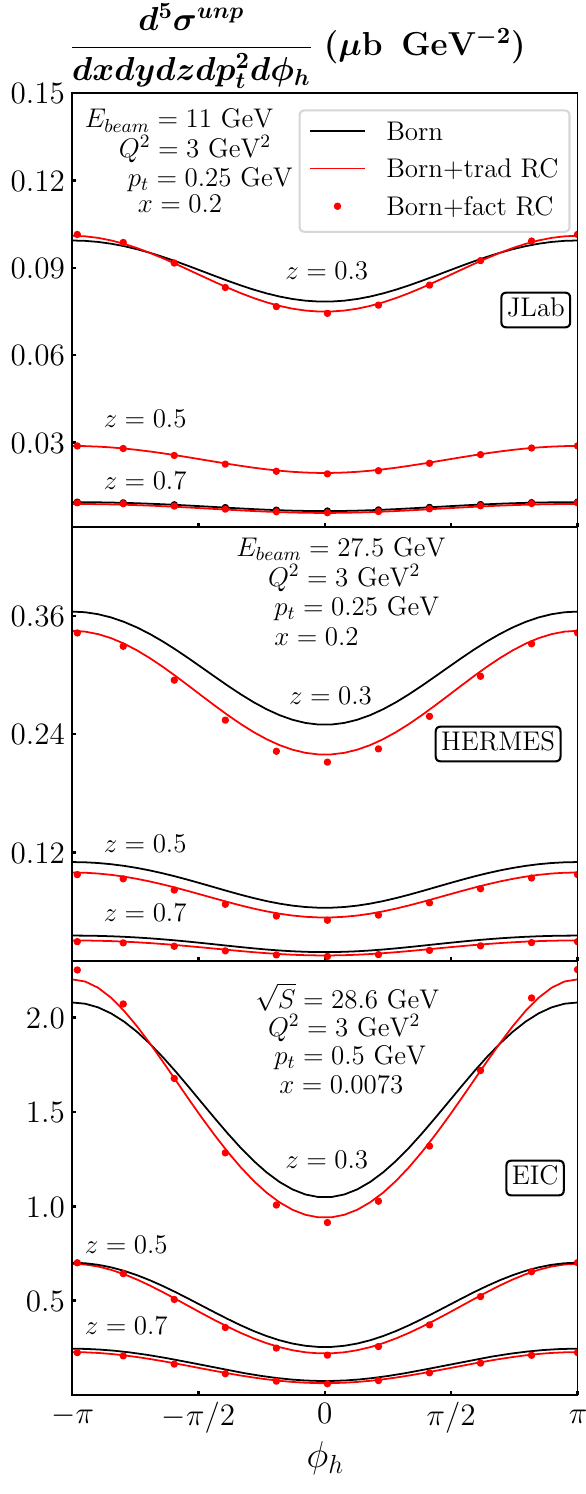}}
\vspace{-3.5mm}
\caption{
Dependence on the azimuthal angle $\phi_h$ of the SIDIS unpolarized Born (black lines) and observed (radiation-included) cross sections for semi-inclusive $\pi^+$ electroproduction at JLab ($E_{\rm beam}=11$~GeV), HERMES ($E_{\rm beam}=27.5$~GeV), and EIC ($\sqrt{S}=28.6$~GeV) kinematics, in the traditional (red lines) and factorized (red dots) approaches.
}
\label{fig:fig_unpoll_xs}
\end{figure*}

\begin{figure*}[p]\centering
\scalebox{0.85}{\includegraphics[width=1.1\textwidth, height=1.05\textheight]{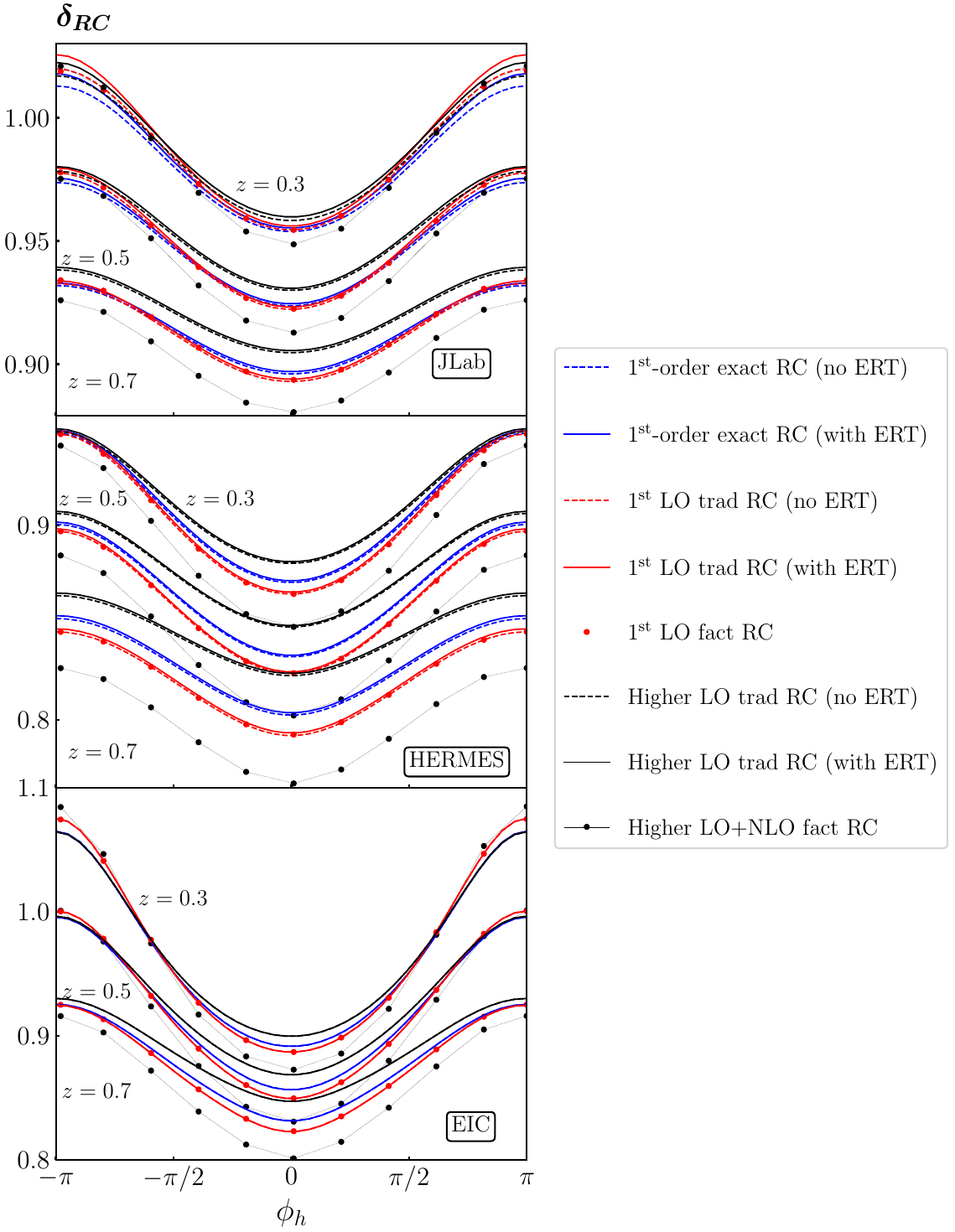}}
\vspace{-1.5mm}
\caption{
$\phi_h$ dependence of the RC factor for semi-inclusive $\pi^+$ electroproduction at the same JLab, HERMES, and EIC kinematic conditions as in Fig.~\ref{fig:fig_unpoll_xs}. The results for $\delta_{\rm RC}$ in the traditional approach computed with the complete RCs (black lines) are compared with the $1^{\rm st}$-order exact (blue lines) and leading-log approximation (red lines), including the ERT (solid lines) and without the ERT (dashed lines). The results for $\delta_{\rm RC}$ in the factorized approach with the complete RCs (black points connected by thin black lines) are compared with the leading-log approximation (red points).
}
\label{fig:fig_unpoll_drc}
\end{figure*}

\begin{figure*}[hbtp]\centering
\scalebox{0.85}{\includegraphics{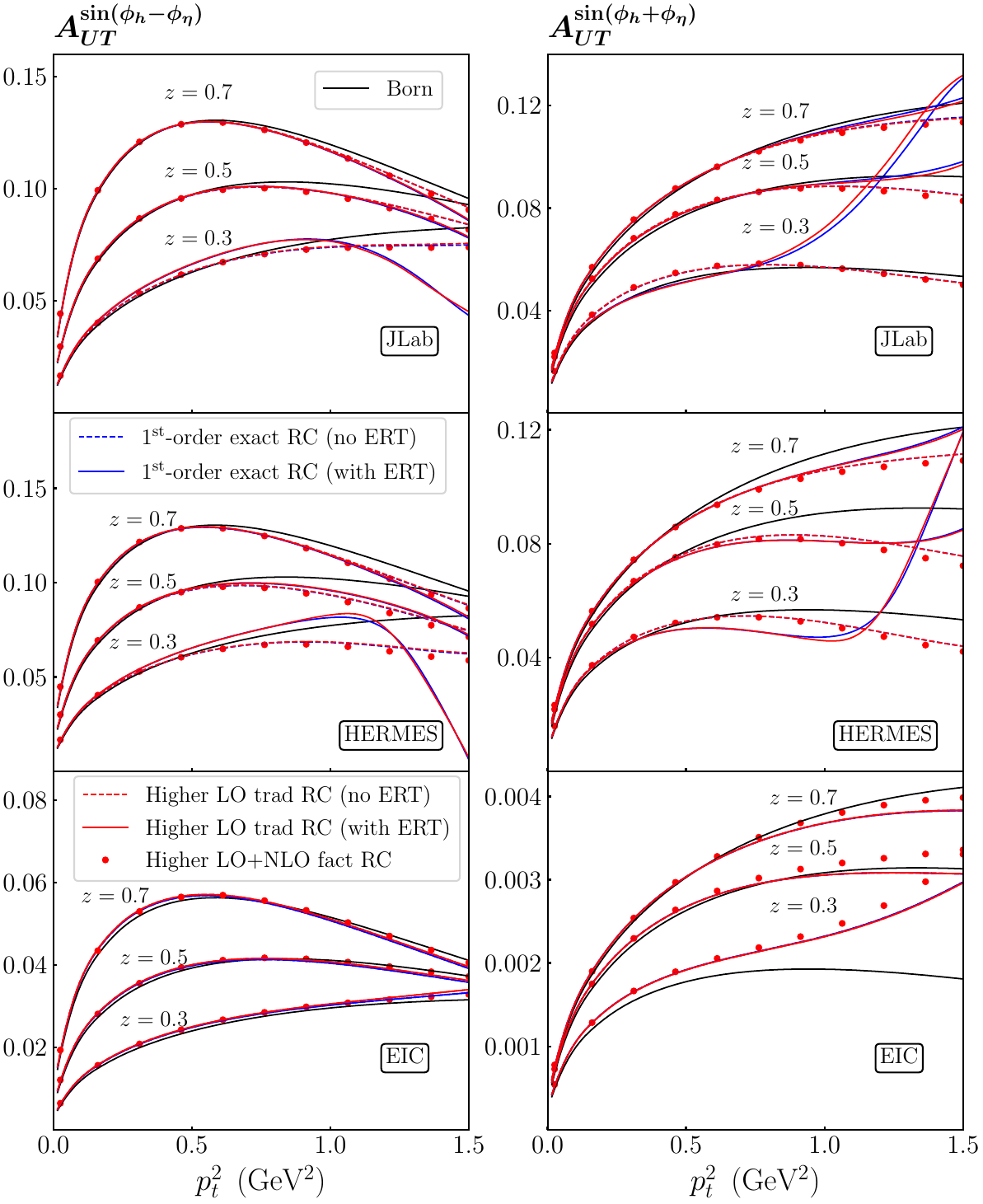}}
\vspace{-2.5mm}
\caption{
$p_t$ dependence of the Sivers (left) and Collins (right) SSAs for semi-inclusive $\pi^+$ electroproduction for the same JLab, HERMES, and EIC kinematics as in Fig.~\ref{fig:fig_unpoll_xs}. 
For both asymmetries, the black, blue, and red solid curves describe the Born contribution, Born contribution with exact RCs, and Born contribution with RCs in the leading-log approximation at 
the lowest order in $\alpha$, respectively. The solid and dashed curves are obtained in the traditional approach with and without inclusion of the ERT, respectively. The red points are obtained 
in the leading-log approximation in the factorized approach.
}
\label{fig:fig_asym}
\end{figure*}

The results of the numerical comparisons are illustrated in Figs.~\ref{fig:fig_unpoll_xs}--\ref{fig:fig_asym}.
Since the same parametrizations of the semi-inclusive structure functions are employed in both the traditional and factorized approaches, all observables are identical at the Born level. 
For the notations used in the figures:
\begin{itemize}

\item[$\bullet$] 
``$1^{\rm st}$-order exact RC" refers to the first‑order (with respect to $\alpha$) RC contribution, given in the traditional approach by Eq.~(\ref{eq:siga_obs}), which is calculated exactly rather than in any approximation.

\item[$\bullet$] 
``$1^{\rm st}$ LO trad RC" and ``$1^{\rm st}$ LO fact RC" refer to the traditional [Eqs.~(\ref{xs_contin}) and (\ref{sllex1})] and factorized [Eq.~(\ref{eq:xs_alpha_factor})] first leading‑order RC contributions with respect to $\alpha$ ($\sim \alpha L$).

\item[$\bullet$] 
``Higher LO trad RC" refers to the all‑order resummation of leading logarithms (full order of $(\alpha L)^{n}$) [Eqs.~(\ref{inll}) and (\ref{exll})], while
``Higher-order LO+NLO fact RC" refers to the all-order resummation of leading logarithms (complete tower of terms of the form of $\alpha^n L^n$), together with a partial resummation of next-to-leading logarithms (subset of terms of the form of $\alpha^n L^{n-1}$) [Eq.~(\ref{RC_x_fac})].

\end{itemize}

The unpolarized Born and observed SIDIS cross sections are presented in Fig.~\ref{fig:fig_unpoll_xs} as a function of $\phi_h$ for selected kinematics at JLab, HERMES, and the future EIC (for a lower 
center-of-mass energy). The radiation-included cross section in the traditional approach is calculated exactly from the $1^{\rm st}$-order expression in Eq.~(\ref{eq:siga_obs}), along with the higher-order 
RCs calculated using the electron structure functions in Eqs.~(\ref{inll}) and (\ref{exll}). 
The corresponding cross section calculated in the factorized approach with LO+NLO RCs is obtained from Eq.~(\ref{RC_x_fac}).
The results are essentially consistent between the two approaches, given the kinematical conditions, with the cross section relative difference, defined as
\begin{displaymath}
\mbox{Relative difference} 
= \frac{\sigma^{\rm obs}_{\rm trad}}
       {\sigma^{\rm obs}_{\rm fact}} - 1,
\end{displaymath}
varying from $\approx 1\%$ up to a few $\%$ at most for $\phi_{h} = 0$, as Table~\ref{tab:ratio} illustrates.

\setlength{\tabcolsep}{4pt}
\vskip 0.0truecm
\begin{table}[t] 
\caption{Relative differences (in \%) between cross sections in the traditional and factorized approaches from Fig.~\ref{fig:fig_unpoll_xs} at JLab, HERMES, and EIC kinematics, for several values of $\phi_h$ and $z$.}
\label{tab:ratio}
\centering
\begin{tabular}{c|c|c|c}
\hline
Rel. diff. (\%) & JLab & HERMES & EIC \\
\hline
$\phi_{h}$ & $0,~\pi/2, ~\pi$ & $0,~\pi/2, ~\pi$ & $0,~\pi/2, ~\pi$ \\
\hline
\hline
$z = 0.3$ & 0.8,~0.1,~$-0.50$ & 3.6,~2.2,~0.6 & 2.9,~0.8,~$-2.3$ \\ 
$z = 0.5$ & 1.6,~0.8,~$-0.08$ & 5.4,~3.9,~2.2 & 4.3,~2.0,~$-0.9$ \\
$z = 0.7$ & 2.3,~1.8,~~~\,0.80 & 6.8,~5.8,~4.2 & 5.3,~3.4,~~~1.2 \\ 
\hline
\end{tabular}
\end{table}

The corresponding $\delta_{\rm RC}$ factor is shown in Fig.~\ref{fig:fig_unpoll_drc} as a function of $\phi_{h}$ for the same three kinematic conditions as in Fig.~\ref{fig:fig_unpoll_xs}.
The results for the traditional approach are shown with and without the ERT contribution.
The largest differences observed are those between the higher LO traditional RC and higher LO+NLO factorized RC results, which partially reflects the differences between the traditional and factorized calculations shown in Fig.~\ref{fig:fig_unpoll_xs}.
The main source of the discrepancy between the two approaches originates from the terms containing $\log(1-x)$ in the initial conditions of the LDFs and LFFs (see Eq.~\eqref{eq:NLOLDFee}), which appear in the factorized approach but are absent in the traditional approach. 
We refer to Appendix~\hyperref[sec:appB]{B} for the derivation of the specific $\log(1-x)$ term in the LDFs.
Note that the LDFs and LFFs in the factorized approach are in general nonperturbative quantities and can in principle be extracted from experimental data. 
For our numerical comparisons, however, since experimentally extracted LDFs and LFFs are not yet available, we instead use the perturbative approximations given in Eqs.~(\ref{eq:NLOLDF}).
These perturbative results may differ from the full LDFs and LFFs \cite{Cammarota:2025jyr,Qiu:2026fed}.

The effects of RCs on the Sivers and Collins transverse SSAs are shown in Fig.~\ref{fig:fig_asym}. The main observations from these results include:
\begin{itemize}

\item[\textbf{(a)}]
The complete RCs from both approaches and the 1$^{\mathrm{st}}$-exact RC from the traditional approach are almost identical among each other.

\item[\textbf{(b)}]
Both approaches provide quite similar results for the asymmetries, although the largest difference is observed for the Collins SSA at EIC kinematics, which at large $p_t^2 = 1.4\,{\rm GeV^2}$ can reach up to (11.3\%, 8.5\%, 3.0\%) at $z = (0.3, 0.5, 0.7)$.

\item[\textbf{(c)}]
The effects of the ERT are conspicuously large in the high-$p_t$ region (in particular, for $z = 0.3$) at JLab and HERMES kinematics, but for the EIC chosen kinematics the contribution of the radiative tail from the exclusive peak is negligible.

\item[\textbf{(d)}]
The RC effects cause quite large discrepancies between the Born-level and radiation-included SSAs, especially again in the high-$p_t$ region and for $z = 0.3$ (except for the Sivers SSA at the EIC). 
The asymmetry relative difference at $p_t^2 = 1.4\,{\rm GeV^2}$ is (65\%, 78\%) at [JLab, HERMES] for the Sivers asymmetry, and (226\%, 178\%, 152\%) at [JLab, HERMES, EIC] kinematics for the Collins asymmetry.

\end{itemize}

We note that if the contribution of a certain effect is similar for both spin-dependent and spin-averaged parts, then this effect will not contribute to the lowest-order correction to an asymmetry.
These mutual cancellations is a cause of occurrence of conclusions in the observations {\bf (a)} and {\bf (b)}. 
The findings {\bf (c)} and {\bf (d)} could be partially explained if we kept in mind the role of the small-$W$ threshold region in the integration (the most important in the ERT calculations \cite{2009}), whose fraction is negligible at the EIC and reasonably large in fixed-target experiments. 

More specific contributions to the RCs and numerical details for the three observables discussed in this section are given in Tables~\ref{tab:unpol}, \ref{tab:sivers}, and \ref{tab:collins} for the JLab kinematics shown in Fig.~\ref{fig:fig_unpoll_xs}. 
The displayed results provide a more detailed picture for the comparison between the two approaches discussed in this work. 
The effect of the $\log(1-x)$ term in the NLO correction is visible especially at [$z=0.7$, $\phi_h=0$] for $\delta_{RC}$ in Table~\ref{tab:unpol}, when $\delta_{RC}$ is between 0.8929 (fifth row) and 0.9020 (eighth row), and when the contributions of the $1^{\rm st}$-order exact RC, plus LO, ERT, and higher LO RCs, are partially and completely combined together. 
Meanwhile, the RC factor $\delta_{RC}$ is 0.8758 (second row) for the lowest order NLO and 0.8805 (ninth row) for the total contribution, when all the NLO contributions are partially summed. \\

\setlength{\tabcolsep}{4.5pt}
\begin{table*}[t] 
\caption{Various contributions to the $\delta_{RC}$ factor of the unplolarized cross section from Eq.~(\ref{eq:delta_RC_factor}) determined at typical JLab kinematics corresponding to Fig.~\ref{fig:fig_unpoll_drc}, for a range of values of [$z$, $\phi_{h}$].}
\label{tab:unpol}
\centering
\begin{tabular}{l|ccc|ccc|ccc}
\hline\hline
\hspace*{2.2cm} [$z$, $\phi_{h}$] 
& [$0.3, 0$] & [$0.3, \pi/2$] & [$0.3, \pi$]
& [$0.5, 0$] & $[0.5, \pi/2$] & [$0.5, \pi$]
& [$0.7, 0$] & [$0.7, \pi/2$] & [$0.7, \pi$] \\ 
 \hline
 \hline
$1^{\rm st}$ LO from Eq.~(\ref{eq:xs_alpha_factor}) 
& 0.9544 & 0.9830 & 1.0188 
& 0.9224 & 0.9481 & 0.9778 
& 0.8936 & 0.9131 & 0.9341 \\ 
\hline
$1^{\rm st}$ LO+NLO from Eq.~(\ref{eq:xs_alpha_factor2}) 
& 0.9499 & 0.9837 & 1.0252 
& 0.9114 & 0.9429 & 0.9787 
& 0.8758 & 0.9007 & 0.9269 \\
\hline
$1^{\rm st}$-order exact formula from Eq.~(\ref{eq:siga_obs2}) 
& 0.9538 & 0.9802 & 1.0129 
& 0.9236 & 0.9471 & 0.9737 
& 0.8960 & 0.9136 & 0.9319 \\ 
\hline
$1^{\rm st}$-order exact formula from Eq.~(\ref{eq:siga_obs}) 
&  &  &  &  &  &  &  &  &  \\
including the ERT 
& 0.9552 & 0.9822 & 1.0179 
& 0.9244 & 0.9481 & 0.9754
& 0.8969 & 0.9145 & 0.9329 \\
\hline
$1^{\rm st}$ LO from Eq.~(\ref{xs_contin}) 
& 0.9544 & 0.9834 & 1.0198 
& 0.9223 & 0.9481 & 0.9776 
& 0.8929 & 0.9123 & 0.9327 \\ 
\hline
$1^{\rm st}$ LO from Eq.~(\ref{xs_contin}) plus 
&  &  &  &  &  &  &  &  &  \\
the ERT from Eq.~(\ref{sllex1}) 
& 0.9560 & 0.9857 & 1.0256
& 0.9231 & 0.9492 & 0.9796 
& 0.8939 & 0.9133 & 0.9338 \\
\hline
the above contribution (third line) 
&  &  &  &  &  &  &  &  &  \\
with higher LO contributions
&  &  &  &  &  &  &  &  &  \\
based on Eq.~(\ref{inll})
& 0.9547 & 0.9798 & 1.0113
& 0.9265 & 0.9481 & 0.9730 
& 0.9011 & 0.9168 & 0.9333 \\
\hline
the above contribution (fourth line)
&  &  &  &  &  &  &  &  &  \\
with higher LO contributions 
&  &  &  &  &  &  &  &  &  \\
based on Eqs.~(\ref{inll}) \& (\ref{exll}) 
& 0.9562 & 0.9819 & 1.0163
& 0.9273 & 0.9492 & 0.9747
& 0.9020 & 0.9177 & 0.9343 \\
\hline
full contribution from Eq.~(\ref{RC_x_fac})
& 0.9486 & 0.9807 & 1.0210 
& 0.9127 & 0.9417 & 0.9752 
& 0.8805 & 0.9025 & 0.9259 \\
\hline\hline
\end{tabular}
\end{table*}

\setlength{\tabcolsep}{4.5pt}
\begin{table*}[t] 
\vskip -0.1truecm
\caption{Various contributions to the Sivers asymmetry $A_{UT}^{\sin(\phi_h-\phi_\eta)}$ determined at the typical JLab kinematics corresponding to Fig.~\ref{fig:fig_asym}, for a range of values of [$z$, $p_t^2$~(GeV$^2$)].}
\label{tab:sivers}
\centering
\begin{tabular}{l|ccc|ccc|ccc}
\hline\hline
\hspace*{1.8cm} [$z$, $p_{t}^{2}~{\rm (GeV^{2})}$] 
& [0.3,0.1] & \!\![0.3,0.5]\!\! & [0.3,1.0] 
& [0.5,0.1] & \!\![0.5,0.5]\!\! & [0.5,1.0] 
& [0.7,0.1] & \!\![0.7,0.5]\!\! & [0.7,1.0] \\ 
\hline
\hline
Born contribution from Eq.~(\ref{scasym1}) & 0.0314 & 0.0628 & 0.0775 & 0.0554 & 0.0975 & 0.1024 & 0.0818 & 0.1299 & 0.1199 \\ 
\hline
$1^{\rm st}$ LO from Eq.~(\ref{eq:xs_alpha_factor}) & 0.0323 & 0.0633 & 0.0740 & 0.0563 & 0.0971 & 0.0976 & 0.0826 & 0.1295 & 0.1170 \\ 
\hline
$1^{\rm st}$ LO+NLO from Eq.~(\ref{eq:xs_alpha_factor2}) & 0.0324 & 0.0634 & 0.0738 & 0.0565 & 0.0971 & 0.0970 & 0.0829 & 0.1295 & 0.1164 \\
\hline
$1^{\rm st}$-order exact formula from Eq.~(\ref{eq:siga_obs2}) & 0.0322 & 0.0632 & 0.0739 & 0.0562 & 0.0972 & 0.0981 & 0.0825 & 0.1295 & 0.1173  \\ 
\hline
$1^{\rm st}$-order exact formula from Eq.~(\ref{eq:siga_obs}) &  &  &  &  &  &  &  &  &  \\
including the ERT & 0.0329 & 0.0663 & 0.0770 & 0.0562 & 0.0974 & 0.0981 & 0.0825 & 0.1295 & 0.1166 \\
\hline
$1^{\rm st}$ LO from Eq.~(\ref{xs_contin}) & 0.0323 & 0.0634 & 0.0742 & 0.0563 & 0.0973 & 0.0982 & 0.0826 & 0.1296 & 0.1173 \\ 
\hline
$1^{\rm st}$ LO from Eq.~(\ref{xs_contin}) plus &  &  &  &  &  &  &  &  &  \\
the ERT from Eq.~(\ref{sllex1}) & 0.0330 & 0.0665 & 0.0766 & 0.0563 & 0.0975 & 0.0982 & 0.0826 & 0.1296 & 0.1168 \\
\hline
higher LO contributions &  &  &  &  &  &  &  &  &  \\
based on Eq.~(\ref{inll}) & 0.0323 & 0.0634 & 0.0742 & 0.0563 & 0.0973 & 0.0982 & 0.0826 & 0.1296 & 0.1173 \\ 
\hline
higher LO contributions &  &  &  &  &  &  &  &  &  \\
based on Eqs.~(\ref{inll}) \& (\ref{exll}) & 0.0330 & 0.0665 & 0.0766 & 0.0563 & 0.0975 & 0.0982 & 0.0826 & 0.1296 & 0.1168 \\
\hline
full contribution from Eq.~(\ref{RC_x_fac}) & 0.0324 & 0.0633 & 0.0735 & 0.0564 & 0.0971 & 0.0971 & 0.0827 & 0.1295 & 0.1167 \\
\hline\hline
\end{tabular}
\end{table*}

\setlength{\tabcolsep}{4.5pt}
\begin{table*}[t] 
\vskip -0.1truecm
\caption{As in Table~\ref{tab:sivers}, but for the Collins asymmetry $A_{UT}^{\sin(\phi_h+\phi_\eta)}$.}
\label{tab:collins}
\centering
\begin{tabular}{l|ccc|ccc|ccc}
\hline\hline
\hspace*{1.8cm} [$z$, $p_{t}^{2}~{\rm (GeV^{2})}$] 
& [0.3,0.1] & \!\![0.3,0.5]\!\! & [0.3,1.0] 
& [0.5,0.1] & \!\![0.5,0.5]\!\! & [0.5,1.0] 
& [0.7,0.1] & \!\![0.7,0.5]\!\! & [0.7,1.0] \\ 
\hline
\hline
Born contribution from Eq.~(\ref{scasym2}) & 0.0290 & 0.0524 & 0.0568 & 0.0403 & 0.0775 & 0.0908 & 0.0440 & 0.0892 & 0.1116 \\ 
\hline
$1^{\rm st}$ LO from Eq.~(\ref{eq:xs_alpha_factor}) & 0.0311 & 0.0552 & 0.0570 & 0.0422 & 0.0790 & 0.0881 & 0.0456 & 0.0900 & 0.1086 \\ 
\hline
$1^{\rm st}$ LO+NLO from Eq.~(\ref{eq:xs_alpha_factor2}) & 0.0314 & 0.0557 & 0.0573 & 0.0426 & 0.0794 & 0.0880 & 0.0460 & 0.0902 & 0.1082 \\
\hline
$1^{\rm st}$-order exact formula from Eq.~(\ref{eq:siga_obs2}) & 0.0309 & 0.0550 & 0.0570 & 0.0420 & 0.0789 & 0.0885 & 0.0453 & 0.0898 & 0.1092 \\ 
\hline
$1^{\rm st}$-order exact formula from Eq.~(\ref{eq:siga_obs}) &  &  &  &  &  &  &  &  &  \\
including the ERT & 0.0299 & 0.0513 & 0.0701 & 0.0419 & 0.0785 & 0.0896 & 0.0453 & 0.0898 & 0.1101 \\
\hline
$1^{\rm st}$ LO from Eq.~(\ref{xs_contin}) & 0.0310 & 0.0552 & 0.0569 & 0.0421 & 0.0790 & 0.0884 & 0.0454 & 0.0899 & 0.1090 \\ 
\hline
$1^{\rm st}$ LO from Eq.~(\ref{xs_contin}) plus &  &  &  &  &  &  &  &  &  \\
the ERT from Eq.~(\ref{sllex1}) & 0.0299 & 0.0511 & 0.0724 & 0.0419 & 0.0786& 0.0894 & 0.0454 & 0.0899 & 0.1097 \\
\hline
higher LO contributions &  &  &  &  &  &  &  &  &  \\
based on Eq.~(\ref{inll}) & 0.0310 & 0.0552 & 0.0569 & 0.0421 & 0.0790 & 0.0884 & 0.0454 & 0.0899 & 0.1090 \\ 
\hline
higher LO contributions &  &  &  &  &  &  &  &  &  \\
based on Eqs.~(\ref{inll}) \& (\ref{exll}) & 0.0299 & 0.0511 & 0.0724  & 0.0419 & 0.0786& 0.0894 & 0.0454 & 0.0899 & 0.1097\\
\hline
full contribution from Eq.~(\ref{RC_x_fac}) & 0.0314 & 0.0556 & 0.0570 & 0.0425 & 0.0793 & 0.0878 & 0.0458 & 0.0901 & 0.1083 \\
\hline\hline
\end{tabular}
\end{table*} 

The RC contributions for the SSAs shown in Tables~\ref{tab:sivers} and \ref{tab:collins} demonstrate relatively small effects for both Sivers and Collins asymmetries. 
For example, if the Born Sivers asymmetry contribution at [$z=0.7$, $p_t^2=0.5~\rm GeV^2$] is 0.1299 (first row), then the RC contributions span between 0.1295 and 0.1296, as shown in the corresponding column 
of Table~\ref{tab:sivers}. 
Similarly, the Born Collins asymmetry at these kinematics is 0.0892 (first row), and the RC-corrected asymmetry variations are minor, varying from 0.0898 to 0.0902, as shown in the corresponding column of Table~\ref{tab:collins}.
The only exception from this is the contribution of ERT for the Collins SSA; for instance, the RC-corrected asymmetry at [$z = 0.3$, $p_t^2 = 1.0~\rm GeV^2$] in Table~\ref{tab:collins} with and without ERT is 0.0724 (seventh/ninth rows) and 0.0569 (sixth/eighth rows), respectively.

\clearpage
\section{Discussion \label{eq::dis}}

High-precision SIDIS experiments are planned as part of the 12 GeV program at JLab~\cite{JLAB12, SoLID1, SoLID2, SoLID3} and at the future EIC~\cite{Accardi:2012qut, AbdulKhalek:2021gbh}. 
Data from these experiments will provide a unique opportunity to explore the 3-dimensional partonic structure of the nucleon in momentum space through the study of TMD PDFs. 
At present, RCs in SIDIS processes are one of the major sources of systematic uncertainty in extracting TMD PDFs, and in the present study we have made a critical comparison between the different approaches to RCs that have recently been discussed in the literature.

In the traditional approach to RCs, originally developed by Bardin and Shumeiko \cite{Bardin:1976qa} and subsequently extended to unpolarized and polarized SIDIS processes by Akushevich, Ilyichev {\it et al.}~\cite{2009, Akushevich:2019mbz, Akushevich:2024uhb, Tra_RC_exc}, the core is an exact method for extraction and cancellation of the infrared divergence at lowest order in $\alpha$.
Other robust features include opportunities to
(i) perform a separate calculation of the radiative tail from the exclusive peak, when a radiated photon is accompanied by only one unobserved hadron in the final state; and
(ii) incorporate higher-order effects in $\alpha$ using the method of electron structure functions in its original formulation by Kuraev and Fadin~\cite{kuraev1, kuraev2}. 

The alternative approach is based on the joint QED+QCD factorization methodology~\cite{Liu:2020rvc, Liu:2021jfp}, following earlier work~\cite{Frixione:2019lga, Bertone:2019hks, Frixione:2012wtz, Bertone:2022ktl, Arbuzov:2024tac} on factorization in the presence of QED radiation.
The core of this approach is the factorization of collision-induced QED radiation into universal LDFs and LFFs, in analogy with the treatment of collinear radiation in QCD, leaving infrared-safe, short-distance hard parts that are calculable perturbatively order by order in $\alpha$ and $\alpha_s$~\cite{Cammarota:2025jyr}.
The scale dependence of the LDFs and LFFs is governed by DGLAP-type evolution equations~\cite{Dsp, GLsp, APsp}, which allow resummation of logarithmically enhanced QED contributions. 
The strength of this approach is its systematic and model-independent organization of QED radiative effects, a feature particularly useful in SIDIS, where radiation modifies the effective momentum transfer and distorts the reconstructed photon--nucleon frame. 

The framework for the comparison of predictions provided by the two approaches involves two stages.
First, we isolate the leading-log RCs at lowest order and compare the analytical expressions obtained from both approaches (see Sec.~\ref{eq::RCs_comp_form}).
In the second stage, all the contributions available in each approach are evaluated numerically, and their magnitudes compared with each other and with the total correction (see Sec.~\ref{eq::RCs_comp_obs}).
We note that the isolation of leading-log terms from the results provided by each approach makes the expressions for the RCs more transparent.
Furthermore, there are at least three situations when such leading-log approximations are necessary and adequate:
(i) when exact calculations are not possible, such as for exact contribution-by-contribution calculations at higher orders;
(ii) when one makes theoretical formulations, manipulations and illustrations to clarify the physics of the processes under consideration;
(iii) when one contrasts different computer codes for RC calculations, since the extraction of leading terms to be compared (from those codes) 
often is the most effective way to identify sources of disagreement.

Within the traditional approach, the lowest-order RCs are calculated exactly, or what is almost equivalent to the ultrarelativistic approximation, in which terms proportional to the electron mass squared are omitted in the final expressions. 
Since the lowest-order cross section can be written in the form $\log(Q^2/m^2) A + B + \mathcal{O}(m^2/Q^2)$, where $A$ and $B$ do not depend on the electron mass, both the $A$ (LO) and $B$ (NLO) terms are kept in the final expressions for RCs. 
In the factorized approach, the same terms are kept when we consider the LO and NLO RCs, excluding those terms that cannot be classified as collinear.
However, keeping only the collinear terms does not make both approaches equivalent at lowest order because an additional assumption is made in the factorized approach.
Namely, hadronic structure functions are determined in the collinear peaks that result in a final expression in the form of a one-dimensional integral over the photon energy (or $\xi$ and/or $\zeta$), while in the traditional approach three-dimensional integration is required for calculating the lowest-order RCs [see Eq.~\eqref{srfin}]. 
Although determining the structure functions in the collinear peaks requires the additional assumption, this is not necessarily a limitation of the factorized approach, and can in fact be considered a strength of the LO/NLO formulas when the hadronic structure functions are not well known in the broad kinematical region required for the three-dimensional integration in the traditional approach.

The numerical comparisons between the two approaches studied here depend on the perturbative LDFs and LFFs, while the full universal LDFs and LFFs can contain also nonperturbative contributions, and can in principle be extracted from data~\cite{Cammarota:2025jyr, Qiu:2026fed}. 
In practice, we find that the $\log(1-x)$ term in the initial conditions of the LDFs and LFFs in Eqs.~\eqref{eq:NLOLDF} has a sizable numerical impact. 
However, as discussed in Appendix~\hyperref[sec:appB]{B} and in Ref.~\cite{Frixione:2012wtz}, this term is factorization-scheme dependent. 
Generally, the LDFs and LFFs can be written as a sum of a scheme-independent collinear-logarithmic contribution and a scheme-dependent contribution. 
In the $\overline{\mathrm{MS}}$ scheme, the $\log(1-x)$ term is part of the latter, whereas it may be absent in a different factorization scheme. 
Its contribution to a scheme-independent physical observable can only be determined after combining the LDFs and LFFs with the hard part in Eq.~\eqref{RC_x_fac}.
Since the $\mathcal{O}(\alpha^3)$ hard part for SIDIS in the factorized approach is not yet available, the cancellation of factorization-scheme dependence currently cannot be implemented. 
Consequently, scheme-independent contributions of LDFs and LFFs to the physical observables cannot be determined unambiguously at the present perturbative accuracy, and the numerically large contributions of the term $\log(1-x)$ must be interpreted with caution.

In addition, we find that the contribution of the radiative tail from the exclusive peak can be extremely important in certain kinematical regions, such as at high $p_t$ and low $z$, for polarized scattering.
In the traditional approach, the ERT is calculated separately from the calculation of the RCs to the continuous spectrum (namely, the multiple production of hadrons), allowing for reduction of one integration over the photon phase space. 
This avoids having to extrapolate the semi-inclusive structure functions to cover exclusive processes.
We note that the ERT could also be analyzed in the factorized approach to extract, quantitatively describe, and address the structure of collinear radiation.

At EIC energies, $Z$-boson exchange and $\gamma Z$ interference may become non-negligible, especially for SIDIS observables at high $Q^2$ and for parity-violating observables in general.
In this case, the electroweak (EW) scale EIC physics requires incorporation of the EW sector into both RC approaches: QED $\to$ QED + EW for the traditional approach, and QED + QCD $\to$ QED + EW + QCD for the factorized approach.
The traditional approach can be further developed by extending the analysis of one-loop EW corrections to polarized DIS~\cite{Akushevich:1995vw} to the case of SIDIS. 
For the factorized formalism, the hard part $\hat{\sigma}$ [see Eqs.~(\ref{RC_x_fac}) and (\ref{eq:fac_born})] should be extended to include EW neutral current SIDIS kernels, including resummation of EW logarithmic effects. 
An important component in these developments will be exploring the renormalization-scheme dependence \cite{Bohm:1986rj, Denner:1991kt, Dittmaier:2026nae}, which is less understood for EW physics than for QED.

The hadronic bremsstrahlung and the TPE (or box) diagrams were not considered in our current work for several reasons. 
First, we included only the model-independent parts of the RCs, which can be calculated analytically without any assumptions about the hadronic structure described by the eighteen SIDIS structure functions in Eq.~(\ref{born_xsection2}). 
Second, the model-independent contribution is numerically the largest one, since the hard photon radiation from the proton is expected to be suppressed by the respective large logarithms, $\log(Q^2/M^2) \ll \log(Q^2/m^2)$. 
The third reason is related to avoiding possible double counting of the model-dependent effects. 
Numerical estimates of RCs usually require specific models of the hadronic structure functions.
These structure functions have typically been extracted from experimental data using some RC prescription, which is not necessarily the same as that used in subsequent RC calculations, thereby leading to potential inconsistencies. 
In the factorized QED+QCD approach, the hadronic structure information is, in principle, extracted at the same time as the radiative effects are applied, thereby avoiding any double counting or inconsistency. 

Numerical analysis of the SSAs in Eqs.~(\ref{eq:AUT}) revealed a strong effect of RC‑induced azimuthal mixing, or contamination, of the Collins asymmetry generated by the Sivers structure function, especially at EIC kinematics. 
The integration over $\phi_h$ and $\phi_\eta$ at the Born level gives rise to the asymmetries in Eqs.~(\ref{scasym}), in which only the Sivers (Collins) structure function contributes to the numerator of the Sivers (Collins) asymmetry.
This is a direct mathematical implication of the angular dependence in the Born cross section. 
However, the cross section has a complex angular dependence when radiative effects are included, and does not reduce to the Born cross section structure in Eq.~(\ref{born_xsection2}). 
The integration in Eqs.~(\ref{eq:AUT}) results in contributions of other spin-dependent structure functions to the Sivers and Collins asymmetries, and this effect cannot be ignored in data analyses of modern high-precision SIDIS experiments, especially at collider kinematics.
A detailed investigation of the effect of the RC‑induced contamination (or azimuthal mixing) in spin asymmetries will be the subject of a future analysis.

\section{Outlook \label{eq::sum}}

The main focus of the analysis in this paper has been on analytical and numerical comparisons of the traditional and factorized approaches to calculating QED RCs in SIDIS, in view of identifying new avenues for further developments both in the SIDIS RC theory and its applications to experimental data analyses. 
We identified the contributions to the SIDIS cross sections that are identical in both approaches, and extracted terms responsible for differences between the predictions from the two frameworks observed in numerical calculations. 
In particular, we demonstrated that the LO RCs coincide exactly (analytically) in both approaches, and that the LO corrections summed to all orders are also identical. 
The main differences were found to arise from terms containing $\log(1-x)$, which are present in the factorized approach (even at lowest order in $\alpha$) but are absent in the traditional approach. 
We demonstrated that the extraction and separate calculation of the ERT is crucial, and must be considered in standard calculations of SIDIS RCs. 
Both types of differences --- for the NLO term with $\log(1-x)$ and the separate calculation of the ERT --- were quantified numerically at the kinematics of current and future SIDIS experiments.

We firstly implemented the factorized approach for SIDIS processes by using LO+NLO initial conditions for the LDFs and LFFs together with LO DGLAP evolution, which sums the LO QED contributions to all orders. 
In these developments, the factorized approach was extended to higher orders in $\alpha$, using the inverse Mellin transformation and numerical integration of the corresponding integrals, with the integrands obtained in Mellin space. 
This technique expands the conventional schemes of DGLAP solutions based on using analytical iterations when an iterative solution is constructed for the subsequent order of perturbative ansatz.
We note that there are limited applications in the literature with LO+NLO corrections for summation to all orders via the inverse Mellin transformation, which has never been applied to the case of SIDIS.

In the future, a hybrid RC framework could be constructed by employing the ``best'' features of both approaches analyzed in this paper. 
For example, some components of the traditional approach could be applied when more accuracy is required for lowest-order RC calculations, or when a separate consideration of the ERT is needed. 
The factorized approach, on the other hand, could be applied when limited information is available for hadronic structure functions. 
Additionally, to investigate connections with other effects, the traditional and factorized approaches could be applied in parallel to enable both analytical and numerical comparisons.

A hybrid RC framework could provide a foundation for future developments to meet the needs of current and upcoming SIDIS data analyses. 
The ``traditional side" of the hybrid framework could be further developed by performing calculations of the second-order RC using the standard Bardin-Shumeiko approach.
This includes the effects of double bremsstrahlung, one-loop effects in processes with one unobserved hard photon, two-loop effects, and the effects in processes of production of unobserved lepton pairs. 
All these effects are included in the electron structure functions, such that the leading terms have to be extracted and compared to the terms contained in these functions.
Non-leading terms could also be compared with the NLO terms provided by the ``factorized side" of the same hybrid framework. 

Subsequently, the hybrid approach could be generalized to incorporate contributions from TPE and hadronic radiation, as well as effects in the EW sector, which will be especially important at the EIC, as discussed in Sec.~\ref{eq::dis} above. These effects should be investigated separately for the ERT and for the inelastic unobserved hadronic state.
Also, the incorporation of the ERT in the factorized framework needs to be assessed.

A further improvement will need to be the calculation of the $\mathcal{O}(\alpha^3)$ NLO hard part in the factorized framework, which contains contributions that cannot be identified with collinear radiation, and therefore are not absorbed into the LDFs or LFFs.
This would allow one to explicitly examine how the sizable $\log(1-x)$ contribution to the NLO LDF and LFF initial conditions is compensated by the corresponding contributions of the hard part. 
Such a study would clarify the physical impact of these endpoint-sensitive terms, reduce the present scheme ambiguity, and provide a more robust basis for precision phenomenology of QED radiative effects in SIDIS.
Speculatively, the progress could be further achieved if one implemented an exact procedure based on using the Bardin-Shumeiko method for extraction and cancellation of the infrared divergence, or using freedom in selecting a specific scheme to obtain optimal initial conditions, for example, which coincides with the lowest-order exact RC.

Eventually, the hybrid approach that we propose would combine the complementary strengths of the traditional~\cite{Akushevich:2019mbz, Tra_RC_exc, Akushevich:2024uhb} and factorized approaches~\cite{Liu:2020rvc, Liu:2021jfp, Cammarota:2025jyr} to construct a best approximation to RCs in SIDIS in a unified framework, in which 
(i) the order ${\cal O}(\alpha)$ corrections are calculated exactly,
(ii) the exclusive RCs are calculated separately,
(iii) the higher order RCs are computed using leading- and next-to-leading-log contributions, and
(iv) radiation from hadrons and quarks is incorporated along with radiation from leptons.
The future development of the entire hybrid RC framework could also be combined with the state-of-the-art TMD factorization formalisms \cite{Barry:2025glq, delRio:2024vvq} to provide a unified approach for analyzing a large spectrum of TMD observables across all energy scales.

\section*{Acknowledgments}

We are grateful to Bishnu Karki and Duane Byer for their contributions in the initial stage of this project. We also thank Alexei Prokudin, Jianwei Qiu, Nobuo Sato, Jia-Yue Zhang and Zhiwen Zhao for 
helpful comments and discussions.  
The work of I.A. is supported in part by discretionary research funds provided by Duke University.
The work of H.G. is supported by the U.S. Department of Energy under Contract No. DE-FG02-03ER41231. 
The work of S.J. is supported by the U.S. Department of Energy under Contract No.~DE-FG02-03ER41231 and by the Brookhaven National Laboratory's Directed Research and Development (LDRD) program under LDRD project number 21-045S. 
The work of V.K. is supported by the discretionary research funds of Prof. Haiyan Gao provided by Duke University and by startup funds awarded to Prof. Wenliang (Bill) Li by the Department of Physics and Astronomy 
at Mississippi State University. 
The work of W.M. is supported by the U.S. Department of Energy, Office of Science, Office of Nuclear Physics under Contract No.~89243126CSC000213.
The work of A.I. is supported by state funding awarded to the Institute for Nuclear Problems of Belorussian State University. 
The work of Y.L. and T.L. is supported by the National Key R\&D Program of China under Contract No. 2024YFA1611004, by the National Natural Science Foundation of China under Contract No. 12321005 and by the Shandong Province Natural Science Foundation under Contract No. ZFJH202303.

\clearpage
\renewcommand{\theequation}{A\arabic{equation}}
\setcounter{equation}{0}
\setcounter{figure}{0}

\section*{\label{sec:appA} Appendix~A: LDF and LFF in the factorized approach}
\addcontentsline{toc}{section}{Appendix~A: LDF and LFF in the factorized approach}

To solve Eqs.~\eqref{eq:DGLAP} with $i,j=e,\bar{e},\gamma$, we first use the one-loop renormalization group equation of the $\overline{\text{MS}}$ QED coupling, $\dd\alpha(Q^2)/\dd\log Q^2 = -\beta_0 \,\alpha^2(Q^2)$, to obtain
\begin{subequations}
\begin{align}
&\frac{\dd f_{i/e}(\xi,Q^2)}{\dd\log\alpha} 
= - \sum_j \frac{P_{ij}(\xi)}{2\pi\beta_0} \otimes_\xi f_{j/e}(\xi,Q^2), 
\\
&\frac{\dd D_{e/i}(\zeta,Q^2)}{\dd\log\alpha} 
= - \sum_j D_{e/j}(\zeta,Q^2) \otimes_\zeta \frac{P_{ji}(\zeta)}{2\pi\beta_0}.
\end{align}
\end{subequations}
These are then transformed to Mellin space, such that the Mellin convolutions become simple products,
\begin{subequations}
\label{appA:prods}
\begin{align}
&\frac{\dd f_{i/e}^N(Q^2)}{\dd \log\alpha} 
= - \sum_j \frac{P_{ij}^N}{2\pi\beta_0} f_{j/e}^N(Q^2), 
\\
&\frac{\dd D_{e/i}^N(Q^2)}{\dd \log\alpha} 
= - \sum_j D_{e/j}^N(Q^2) \frac{P_{ji}^N}{2\pi\beta_0}.
\end{align}
\end{subequations}
Using $P_{\gamma e}=P_{\gamma\bar{e}}$, $P_{e\gamma}=P_{\bar{e}\gamma}$, and $P_{\bar{e}e}=P_{e\bar{e}}=0$, Eqs.~(\ref{appA:prods}) can be decoupled as
\begin{subequations}
\begin{align}
&\frac{\dd}{\dd\log\alpha} f_n^N(Q^2) 
= - \frac{P_{ee}^N}{2\pi\beta_0} f_n^N(Q^2), 
\\
&\frac{\dd}{\dd\log\alpha} 
\begin{pmatrix} f_s^N(Q^2) \\ f_{\gamma/e}^N(Q^2) \end{pmatrix}
= - \frac{\mathbf{P}_s^N}{2\pi\beta_0} 
\begin{pmatrix} f_s^N(Q^2) \\ f_{\gamma/e}^N(Q^2) \end{pmatrix}, 
\\
&\frac{\dd}{\dd\log\alpha} D_n^N(Q^2) 
= - \frac{P_{ee}^N}{2\pi\beta_0} D_n^N(Q^2), 
\\
&\frac{\dd}{\dd\log\alpha} 
\begin{pmatrix} D_s^N(Q^2) \\ D_{e/\gamma}^N(Q^2) \end{pmatrix} 
= - \frac{\mathbb{P}_s^N}{2\pi\beta_0} 
\begin{pmatrix} D_s^N(Q^2) \\ D_{e/\gamma}^N(Q^2) \end{pmatrix}.
\end{align}
\end{subequations}
These are now first-order linear homogeneous ordinary differential equations with constant coefficients, and their solutions are given by Eqs.~\eqref{eq:LDF_solution}, in which the matrix exponential can be calculated as~\cite{Furmanski:1981cw}
\begin{subequations}
\begin{align}
\label{eq:exp_spacelikeP}
r(Q^2)^{-\mathbf{P}_s^N/(2\pi\beta_0)}
&= r(Q^2)^{-\lambda_+/(2\pi\beta_0)} \frac{\mathbf{P}_s^N-\lambda_-\mathbb{I}}{\lambda_+-\lambda_-} 
\nn
&+ r(Q^2)^{-\lambda_-/(2\pi\beta_0)} \frac{\mathbf{P}_s^N-\lambda_+\mathbb{I}}{\lambda_--\lambda_+}, 
\nn
&
\\
\label{eq:exp_timelikeP}
r(Q^2)^{-\mathbb{P}_s^N/(2\pi\beta_0)}
&= r(Q^2)^{-\lambda_+/(2\pi\beta_0)} \frac{\mathbb{P}_s^N-\lambda_-\mathbb{I}}{\lambda_+-\lambda_-} 
\nn
&+ r(Q^2)^{-\lambda_-/(2\pi\beta_0)} \frac{\mathbb{P}_s^N-\lambda_+\mathbb{I}}{\lambda_--\lambda_+},
\nn
&
\end{align}
\end{subequations}
with $\mathbb{I}$ the $2\times2$ identity matrix, and
\begin{align}
\lambda_\pm \equiv \frac{1}{2}\! \left[ P_{ee}^N+P_{\gamma\gamma}^N \pm \sqrt{(P_{ee}^N-P_{\gamma\gamma}^N)^2+8P_{e\gamma}^NP_{\gamma e}^N}\, \right].
\end{align}

The Mellin moments of the splitting kernels in Eqs.~\eqref{eq:split_factor} are given by \\
\begin{subequations}
\label{eq:split_factor_N}
\begin{align}
\label{eq:split_factor_ee_N}
&P_{ee}^N = \frac{3}{2} - \frac{1}{N} - \frac{1}{N+1} - 2\left[\psi(N)-\psi(1)\right], 
\\
\label{eq:split_factor_egamma_N}
&P_{e\gamma}^N = \frac{N^2+N+2}{N(N+1)(N+2)}, 
\\
\label{eq:split_factor_gammae_N}
&P_{\gamma e}^N = \frac{N^2+N+2}{(N-1)N(N+1)}, 
\\
\label{eq:split_factor_gammagamma_N}
&P_{\gamma\gamma}^N = -\frac{2}{3},
\end{align}
\end{subequations}
where $\psi$ is the digamma function.
The Mellin moments of the LDFs and LFFs at the initial scale $m$ in Eqs.~\eqref{eq:NLOLDF} are given by
\begin{subequations}
\label{eq:NLOLDFee_N}
\begin{align}
&f_{e/e}^N(m^2) 
 = D_{e/e}^N(m^2) \nn
&= 1 + \frac{\alpha(m^2)}{2\pi} 
\Biggl\{ - P_{ee}^N - \frac{1}{6N^2(N+1)^2} 
\biggl[ 12
\nn
&\quad + N(N+1) 
        \Bigl( 36 + 12\gamma_E^2 N (N+1)
\nn
&\quad\quad + (-21+2\pi^2) N (N+1) + 12\gamma_E (1+2N) 
        \Bigr)
\nn
&\quad - 12N(N+1) 
        \Bigl( -\psi(N) \bigl( 1+2N(1+\gamma_E+\gamma_E N) 
\nn
&\quad\quad +N(N+1)\psi(N) \bigr) + N(N+1) \psi_1(N) 
        \Bigr)
\biggr]
\Biggr\},
\\
\label{eq:NLOLDFgamma_N}
&f_{\gamma/e}^N(m^2) = \frac{\alpha(m^2)}{2\pi} 
\biggl\{ - P_{\gamma e}^N + \frac{4}{(N-1)^2}
\nn
&\hspace*{3.1cm} - \frac{4}{N^2} + \frac{2}{(N+1)^2} \biggr\}, 
\\
\label{eq:NLOLFFgamma_N}
&D_{e/\gamma}^N(m^2) = 0, 
\\
\label{eq:NLOLDFothers_N}
&f_{\bar{e}/e}^N(m^2) = D_{e/\bar{e}}^N(m^2) = 0,
\end{align}
\end{subequations}
where $\gamma_E$ is the Euler-Mascheroni constant and $\psi_1$ is the trigamma function.

To calculate the inverse Mellin transform in Eqs.~\eqref{eq:LDFintegral}, \eqref{eq:LFee_solution}, we parametrize the integration contour as \cite{Vogt:2004ns}
\begin{align}
N(r) = \begin{cases}
c + r e^{+i\phi}, & r\geq0, \\
c - r e^{-i\phi}, & r<0,
\end{cases}
\end{align}
which satisfies $N(-r)=N(r)^*$.
Since for a real function $F(x)$ we have $F^{N(-r)}=\big(F^{N(r)}\big)^*$, we can write
\begin{align}\label{eq:InvMellinTransformNum}
\mathcal{M}^{-1}[F^N](x) 
= \frac{1}{\pi} \int\limits_0^{\infty} \dd r \, 
\mathrm{Im}\!\left[ e^{i\phi} x^{-(c+re^{i\phi})} F^{c+re^{i\phi}} \right].
\end{align}
The integral over $r$ is calculated numerically using Gaussian quadrature.
Following Ref.~\cite{Vogt:2004ns}, we choose $c=1.9$, which places the contour to the right of all poles of the relevant $F^N$; and $\phi=3\pi/4$, which introduces an exponential damping factor in the
integrand at large $r$.

\renewcommand{\theequation}{B\arabic{equation}}
\setcounter{equation}{0}
\setcounter{figure}{0}

\section*{\label{sec:appB} Appendix~B: Derivation of the LDF}
\addcontentsline{toc}{section}{Appendix~B: Derivation of the LDF in Eq.~(\ref{eq:NLOLDFee})}

The LDF $f_{e/e}(x,Q^2)$ in Eq.~\eqref{eq:NLOLDFee} can be derived in perturbation theory in QED starting from its operator definition in the light-front gauge \cite{Collins:1981uw},
\begin{align}\label{eq:LDFee_def}
f_{0,e/e}(x)
=&\, \int \frac{\dd b^-}{2\pi} e^{-ixp^+b^-} \nonumber\\
&\times \frac{1}{2}\sum_{s} \langle p,s| \bar{\psi}(b^-) \frac{\gamma^+}{2} \psi(0) |p,s\rangle,
\end{align}
where $p$ and $s$ are the momentum and spin of the initial electron, respectively, and $x$ is the fraction of the initial electron's longitudinal momentum carried by the final electron.
The subscript ``$0$'' on the LDF indicates an unrenormalized distribution.

Inserting the vacuum state $|0\rangle$ into Eq.~\eqref{eq:LDFee_def} allows us to write 
\begin{align}
&\int \frac{\dd b^-}{2\pi} e^{-ixp^+b^-} \frac{1}{2} \sum_{s} \langle p,s| \bar{\psi}(b^-) |0\rangle \frac{\gamma^+}{2} \langle0| \psi(0) |p,s\rangle \nonumber\\
&\quad= \delta(x-1)\, \frac{1}{2 p^+} \sum_{s} \langle p,s| \bar{\psi}(0) |0\rangle \frac{\gamma^+}{2} \langle0| \psi(0) |p,s\rangle \nonumber\\
&\quad \equiv \delta(x-1)\, \qty( 1 + V ),
\end{align}
where $V$ is the virtual contribution of $\mathcal{O}(\alpha)$.
Furthermore, inserting the one-photon state $|k,\lambda\rangle$ into Eq.~\eqref{eq:LDFee_def}, where $k$ and $\lambda$ are the photon momentum and helicity, respectively, yields the real-emission 
contribution at $\mathcal{O}(\alpha)$,
\begin{align}
\label{eq.R_LDF}
&R \equiv \int \frac{\dd b^-}{2\pi} e^{-ixp^+b^-} \int \frac{\dd[d-1]{\bm{k}}}{(2\pi)^{d-1} 2k^0} 
\nonumber\\
& \quad\qquad \times \frac{1}{2} \sum_{s,\lambda} \langle p,s| \bar{\psi}(b^-) |k,\lambda\rangle \frac{\gamma^+}{2} \langle k,\lambda| \psi(0) |p,s\rangle \nonumber\\
&= \int \frac{\dd[d-2]{\bm{k}_T}}{(2\pi)^{d-1} 2k^+} \frac{1}{2} \sum_{s,\lambda} \langle p,s| \bar{\psi}(0) |k,\lambda\rangle \frac{\gamma^+}{2} \langle k,\lambda| \psi(0) |p,s\rangle,
\end{align}
where $\bm{k}_T$ denotes the transverse components of $k$. The right-hand side of Eq.~(\ref{eq.R_LDF}) is written in $d=4-2\epsilon$ dimensions, suitable for dimensional regularization.
At $\mathcal{O}(\alpha)$, the correlation function is
\begin{align}
\langle k,\lambda| \psi(0) |p,s\rangle
= \frac{i(\slashed{p}-\slashed{k}-m)}{(p-k)^2-m^2} i e \slashed{\epsilon}^*(k,\lambda) u(p,s),
\end{align}
and using $k^+=(1-x)p^+$, the propagator denominator can be written as
\begin{align}
(p-k)^2-m^2 
= - 2 p \cdot k 
= - \frac{\bm{k}_T^2+(1-x)^2m^2}{1-x}.
\end{align}
Integrating over $\bm{k}_T$, the $(1-x)^2m^2$ term gives rise to the $\log((1-x)^2m^2)$ term in Eq.~\eqref{eq:NLOLDFee}.
In the light-front gauge, the polarization sum for the photon is given by
\begin{align}
\sum_{\lambda} \epsilon^\mu(k,\lambda) \epsilon^{\nu*}(k,\lambda) 
= -\, g^{\mu\nu} + \frac{k^\mu n^\nu + n^\mu k^\nu}{k^+},
\end{align}
where $n$ is a light-like vector with light-front components $n^+=0$, $n^-=1$, and $\bm{n}_T=0$.
The real-emission contribution can then be written as
\begin{align}
R &= \alpha(\mu_R^2) 
\biggl( \frac{e^{\gamma_E}\mu_R^2}{4\pi} \biggr)^\epsilon 
\int \frac{\dd[d-2]{\bm{k}_T}}{(2\pi)^{d-2}} 
\frac{1}{\bigl[ \bm{k}_T^2+(1-x)^2m^2 \bigr]^2} 
\nonumber\\
&\;\times 2\qty[ \bm{k}_T^2 \qty(\frac{1+x^2}{1-x} - \epsilon(1-x)) + m^2 (1-\epsilon)(1-x)^3 ],
\label{eq.R-factor}
\end{align}
where $\mu_R$ is the renormalization scale in the $\overline{\text{MS}}$ scheme.
The $\bm{k}_T$ integral can be calculated in dimensional regularization by taking $\epsilon \to 0$,
\begin{align}
R
&= \frac{\alpha(\mu_R^2)}{2\pi} 
\big( e^{\gamma_E}\mu_R^2 \big)^\epsilon
\int\limits_0^\infty 
\frac{\dd{\bm{k}_T} \bm{k}_T^{1-2\epsilon}}{\Gamma(1-\epsilon) \bigl[\bm{k}_T^2+(1-x)^2m^2\bigr]^2} 
\nonumber\\
&\;\times 2\qty[ \bm{k}_T^2 \qty(\frac{1+x^2}{1-x} - \epsilon(1-x)) + m^2 (1-\epsilon)(1-x)^3 ] 
\nonumber\\
&=\, \frac{\alpha(\mu_R^2)}{2\pi} \frac{1+x^2}{1-x} \biggl(\frac{e^{\gamma_E}\mu_R^2}{(1-x)^2m^2}\biggr)^\epsilon\, \Gamma(\epsilon)(1-\epsilon) 
\nonumber\\
&=\, \frac{\alpha(\mu_R^2)}{2\pi} \frac{1+x^2}{1-x} \qty( \frac{1}{\epsilon} + \log\frac{\mu_R^2}{(1-x)^2m^2} - 1 ),
\end{align}
where terms of order $\mathcal{O}(\epsilon)$ and higher are omitted in the last line.
Up to $\mathcal{O}(\alpha)$, the LDF $f_{0,e/e}(x)$ can then be written as
\begin{align}
&f_{0,e/e}(x)\, =\, (1 + V)\, \delta(x-1)
\nonumber\\
&\quad + \frac{\alpha(\mu_R^2)}{2\pi} \frac{1+x^2}{1-x} \qty( \frac{1}{\epsilon} + \log\frac{\mu_R^2}{(1-x)^2m^2} - 1 ).
\end{align}
The virtual contribution $V$ can be determined from the number sum rule \cite{Collins:1981uw},
\begin{align}
\int\limits_0^1 \dd x \bigl[ f_{0,e/e}(x) - f_{0,\bar{e}/e}(x) \bigr] = 1.
\end{align}
Since $f_{0,\bar{e}/e}(x)=0$ through $\mathcal{O}(\alpha)$, we arrive at
\begin{align}\label{eq:NLOLDFee_bare}
&
f_{0,e/e}(x)\, =\, \delta(x-1) 
\nonumber\\
&+\, \frac{\alpha(\mu_R^2)}{2\pi} 
\bigg[ 
    \frac{1+x^2}{1-x} 
    \bigg( \frac{1}{\epsilon} + \log\frac{\mu_R^2}{(1-x)^2m^2} - 1 )
    \bigg)
\bigg]_+,
\end{align}
where the plus distribution is defined in Eq.~\eqref{eq:plus_def}.

The LDF $f_{e/e}(x,Q^2)$ in Eq.~\eqref{eq:NLOLDFee} is renormalized in the $\overline{\text{MS}}$ scheme by subtracting the $1/\epsilon$ pole in Eq.~\eqref{eq:NLOLDFee_bare}.
(The additional terms $\ln 4\pi-\gamma_E$ are subtracted through the factor $(e^{\gamma_E}/4\pi)^\epsilon$ in Eq.~(\ref{eq.R-factor}).)
Other renormalization (or factorization) schemes are also possible.
For example, Ref.~\cite{Frixione:2012wtz} introduces the $\Delta$ scheme that subtracts all the $\mathcal{O}(\alpha)$ terms in Eq.~\eqref{eq:NLOLDFee_bare}.

\bibliography{biblography}

\end{document}